\documentclass{emulateapj}
\usepackage{url}
\usepackage{graphicx}
\usepackage{epstopdf}

\begin{document}


\title{The Mass-Radius Relation of Young Stars, I: UScoCTIO 5, An M4.5 Eclipsing Binary in Upper Scorpius Observed By K2}

\author{
Adam L. Kraus\altaffilmark{1},
Ann Marie Cody\altaffilmark{2},
Kevin R. Covey\altaffilmark{3},
Aaron C. Rizzuto\altaffilmark{1},
Andrew W. Mann\altaffilmark{1,4},
Michael J. Ireland\altaffilmark{5}
}

\altaffiltext{1}{Department of Astronomy, The University of Texas at Austin, Austin, TX 78712, USA}
\altaffiltext{2}{NASA Ames Research Center, Moffett Field, CA 94035, USA}
\altaffiltext{3}{Western Washington Univ., 516 High St., Bellingham, WA 98225, USA}
\altaffiltext{4}{Harlan J. Smith Fellow}
\altaffiltext{5}{Research School of Astronomy \& Astrophysics, Australian National University, Canberra, ACT 2611, Australia}

\begin{abstract}

We present the discovery that UScoCTIO 5, a known spectroscopic binary in the Upper Scorpius star-forming region ($P=34$ days, $M_{tot} \sin(i) = 0.64 M_{\odot}$), is an eclipsing system with both primary and secondary eclipses apparent in K2 light curves obtained during Campaign 2. We have simultaneously fit the eclipse profiles from the K2 light curves and the existing RV data to demonstrate that UScoCTIO 5 consists of a pair of nearly identical M4.5 stars with $M_A = 0.329 \pm 0.002 M_{\odot}$, $R_A = 0.834 \pm 0.006 R_{\odot}$, $M_B = 0.317 \pm 0.002 M_{\odot}$, and $R_B = 0.810 \pm 0.006 R_{\odot}$. The radii are broadly consistent with pre-main sequence ages predicted by stellar evolutionary models, but none agree to within the uncertainties. All models predict systematically incorrect masses at the 25--50\% level for the HR diagram position of these mid-M dwarfs, suggesting significant modifications to mass-dependent outcomes of star and planet formation. The form of the discrepancy for most model sets is not that they predict luminosities that are too low, but rather that they predict temperatures that are too high, suggesting that the models do not fully encompass the physics of energy transport (via convection and/or missing opacities) and/or a miscalibration of the SpT-$T_{\rm eff}$ scale. The simplest modification to the models (changing $T_{\rm eff}$ to match observations) would yield an older age for this system, in line with the recently proposed older age of Upper Scorpius ($\tau \sim 11$ Myr).

\end{abstract}


 \begin{figure*}
 \epsscale{0.95}
\includegraphics[scale=0.48]{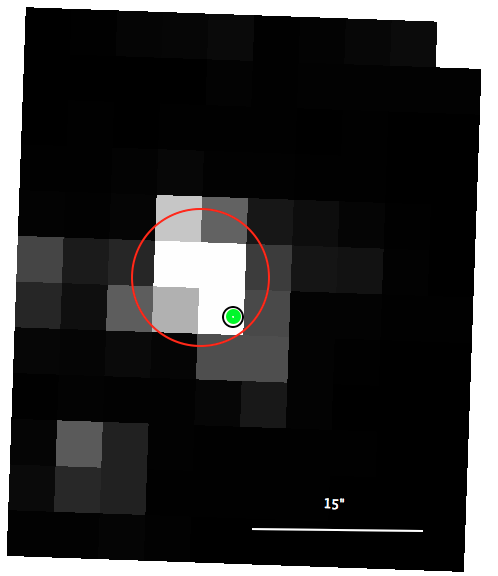}
\includegraphics[scale=0.57,trim=0.0cm -0.1cm 0.0cm 0.0cm]{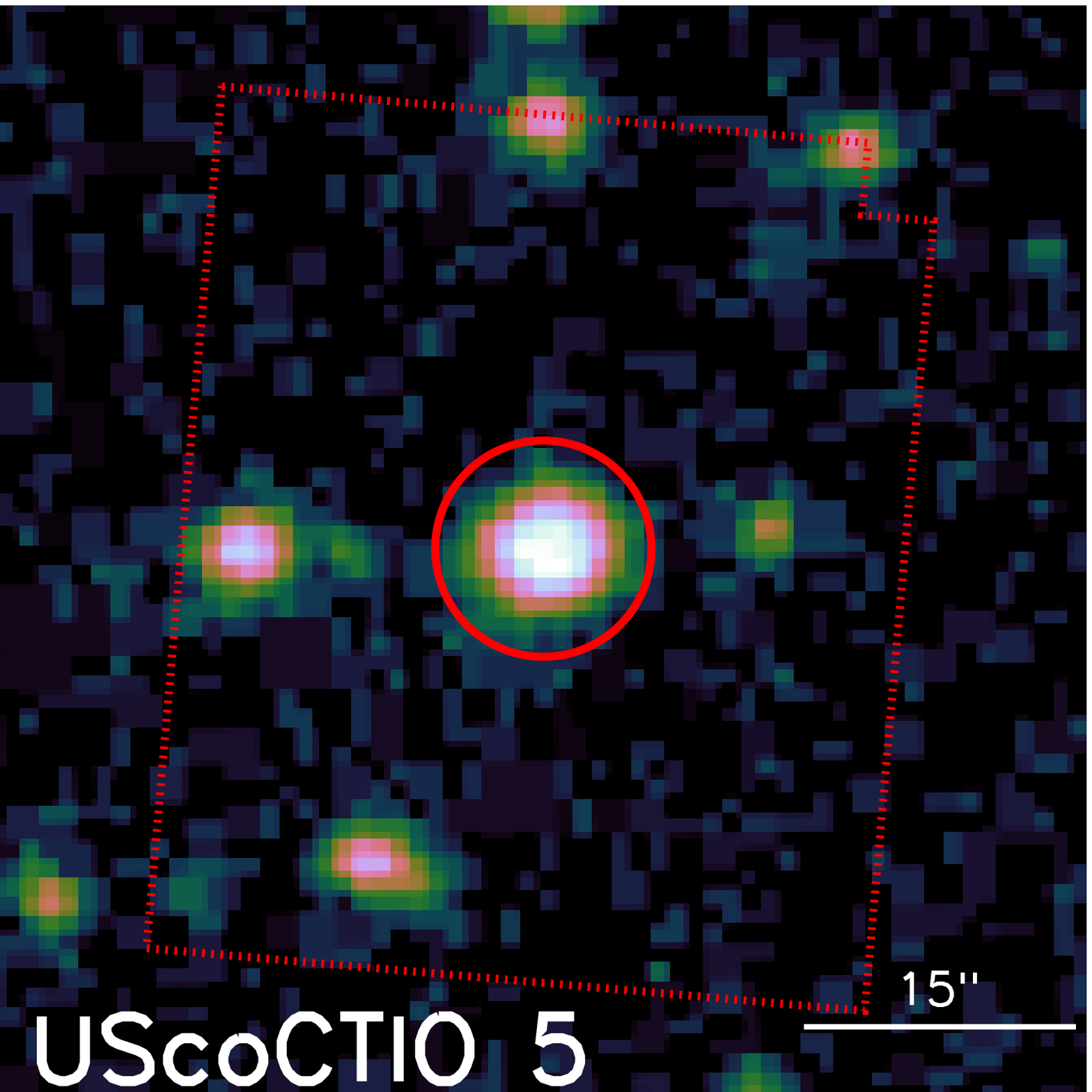} 
 \caption{Left: Postage stamp of the young star UScoCTIO 5 that was downloaded as part of K2's Campaign 2, rotated so that north is up. The red circle shows the 1.5 pixel photometric aperture used in our analysis, after centroiding. The green dot marks the nominal location of the star based on target pixel file header position information. Right: A DSS R-band postage stamp of UScoCTIO 5 (FOV=60\arcsec, North=Up) showing the K2 postage stamp (red dotted box) and our adopted photometric aperture (red circle). The image is shown in a square-root stretch using the CubeHelix color palette (Green 2011). All nearby sources fall outside our photometric aperture. We have previously shown from analysis of 2MASS images that there are no detected sources with a contrast of $\Delta K_s < 3$ at $\rho > 1.5$\arcsec\, and $\Delta K_s < 4$ at $\rho > 2.0$\arcsec \citep{Kraus:2007ve}. UScoCTIO 5 has not been observed with adaptive optics, so there are no limits on closer companions, though the absence of a third set of spectral lines in our spectra suggests that there are no additional objects within $\Delta R < $1--2.} \label{fig:pics}
  \end{figure*}

\section{Introduction}
  
The fundamental properties of stars constitute a bedrock upon which much of astronomy is built, but there remains a paucity of well-characterized stars for calibrating stellar evolutionary models at young ages (1--10 Myr), particularly for low masses ($<$1 $M_{\odot}$). It is a maxim of stellar astrophysics that the properties and lifecycle of a star are largely set by its mass, and hence it is crucial to calibrate the mass predictions of models. Uncertainties in model-derived properties constitute the dominant systematic error for in-situ measurements of the IMF for young populations \citep{Bastian:2010wb}, determinations of star cluster ages (e.g., \citealt{Preibisch:2001gn} versus \citealt{Pecaut:2012dp}), comparison of protoplanetary disk properties to those of mature planetary systems \citep{Andrews:2013uq}, and binary formation studies (e.g., \citealt{Kraus:2011tg,Kraus:2012bh,Duchene:2013fr}. Mass/age ambiguities strongly limit interpretations for directly-imaged gas giant planets in young populations, where the host-star age and companion luminosity establish the planet mass (e.g., \citealt{Lafreniere:2008oy} and \citealt{Ireland:2011fj} versus \citealt{Pecaut:2012dp}, and also \citealt{Carson:2013fk} versus \citealt{Hinkley:2013lr}). 

At present, different model sets predict masses that differ by 50\% for nominally solar-mass pre-main-sequence stars at the same point in the $L$-$T_{\rm eff}$ HR diagram (e.g., \citealt{Hillenbrand:2004bh,Torres:2010lr,Gennaro:2012oq,Stassun:2014lr}). Furthermore, even the temperature scale for converting spectral type to $T_{\rm eff}$ is uncertain for young stars at the level of 100-200 K (\citealt{Luhman:2003pb}; hereafter L03), constituting a systematic uncertainty on HR diagram positions. Stellar evolutionary models can be tested indirectly, such as by placing binary systems (\citealt{White:1999en,Kraus:2009fk,Torres:2013fk}) or even entire stellar populations \citep{Preibisch:2002qt,Naylor:2009dp,Pecaut:2013zr,Chen:2014fk} on the HR diagram. However, these tests are agnostic to stellar mass and age; they demonstrate whether there are any common ages and masses that reproduce the HR diagram positions, but not whether those values are actually correct. Dynamical masses for young visual binaries have cast much light on this issue (\citealt{Boden:2005bc,Schaefer:2012kx,Simon:2013qy}), and are starting to become available in statistically robust samples \citealt{Schaefer:2014uq}; Rizzuto et al., submitted). However, radius measurements are more time-dependent (particularly among M dwarfs that largely evolve down the Hayashi track to the ZAMS) and have remained elusive. Only a small number of young, low-mass eclipsing binaries are known \citep{van-Eyken:2011dq,Morales-Calderon:2012yq}, and with a few well-characterized exceptions \citep{Stassun:2006bf,Irwin:2007vg,Cargile:2008cr,Gomez-Maqueo-Chew:2012dq,Gillen:2014nx}, those systems are faint and remain only coarsely characterized.

The ages of stars remain similarly uncertain, and are even more difficult to measure without appeals to evolutionary models. Traceback simulations have been notoriously contentious (\citealt{Ortega:2002yg,Makarov:2005eu,Mamajek:2014fv}), and HR diagram ages can differ significantly depending on the stellar mass range considered and the assumed star formation history \citep{Preibisch:2002qt,Pecaut:2012dp, Rizzuto:2015fj}. The ongoing debate over the age of the Upper Scorpius OB association and other young populations \citep{Naylor:2009dp,Bell:2013uq,Mamajek:2014fv} demonstrates that systematic errors remain at the level of factors of 2. However, if an evolutionary model reproduces all of the other observable parameters of a star (mass, radius, $T_{\rm eff}$, and $L_{bol}$), then the model-derived age could represent the most robust possible theoretical estimate of the age for that stellar population.

To address these fundamental issues of stellar astrophysics, and to look for planets, we have initiated a search for eclipsing/transiting systems among all of the known and suspected members of Upper Scorpius and Ophiuchus that fell in the footprint observed by K2, the extended Kepler mission \citep{Howell:2014db} during its Campaign 2. These observations comprised 79 continuous days of observing with a 30 minute cadence, with no gaps, and hence are ideal for identifying all eclipsing binaries with periods on the order of this duration or shorter. In total, we proposed observations of 657 confirmed members (GO2052, PI Covey) and 759 likely candidate members (GO2063, PI Kraus) that were optically visible ($K_p < 16$ mag). Given the size of our sample and the known frequency of short-period binary systems in the field \citep{Raghavan:2010sz}, we expect that $\sim$10--15 new eclipsing binaries should be discovered.

The young Upper Scorpius member UScoCTIO 5 (Figure~\ref{fig:pics})\footnote{The USco candidates identified by \citet{Ardila:2000ay} are sometimes abbreviated as ``USco NN''. However, SIMBAD has assigned the name ``USco 5'' to different candidate member by \citet{Sciortino:1998yq}. We caution the reader to not confuse the two sources if pursuing additional followup observations.} was a high-priority target in our program since it is a known spectroscopic binary, first reported and characterized as such by \citet{Reiners:2005wq}. The membership of UScoCTIO 5 in Upper Scorpius was first proposed by \citet{Ardila:2000ay}, based on its HR diagram position, line of sight extinction, and the presence of H$\alpha$ emission; this membership was confirmed by \citet{Reiners:2005wq} based on the presence of lithium absorption that constrains the age to be $\tau < 30$ Myr. \citet{Reiners:2005wq} also detected the presence of two sets of spectral lines in their first high-resolution spectrum, and subsequently obtained enough additional spectroscopic observations to determine the period ($P = 34$ days), the orbital elements, and a minimum mass (modulo the sine of the orbital inclination). Their inferred minimum mass ($M_{tot} \sin(i) = 0.64 \pm 0.02$ $M_{\odot}$) was larger than the mass predicted by evolutionary models, which suggested both that the models could require modifications or additional physics, and that the system could be close to edge-on (and hence could show eclipses).

In this paper, we report that UScoCTIO 5 is indeed oriented to show both primary and secondary eclipses in its K2 light curve. We re-interpret the sum of data available for the system to determine masses and radii for each star, to test different evolutionary model tracks against the empirical constraints of this system, to calculate an empirical constraint on the SpT-$T_{\rm eff}$ conversion for young mid-M stars, and to determine a new semi-empirical age estimate for the Upper Scorpius OB association.

\section{K2 Light Curves}

 \begin{figure*}
 \epsscale{0.90}
 \plotone{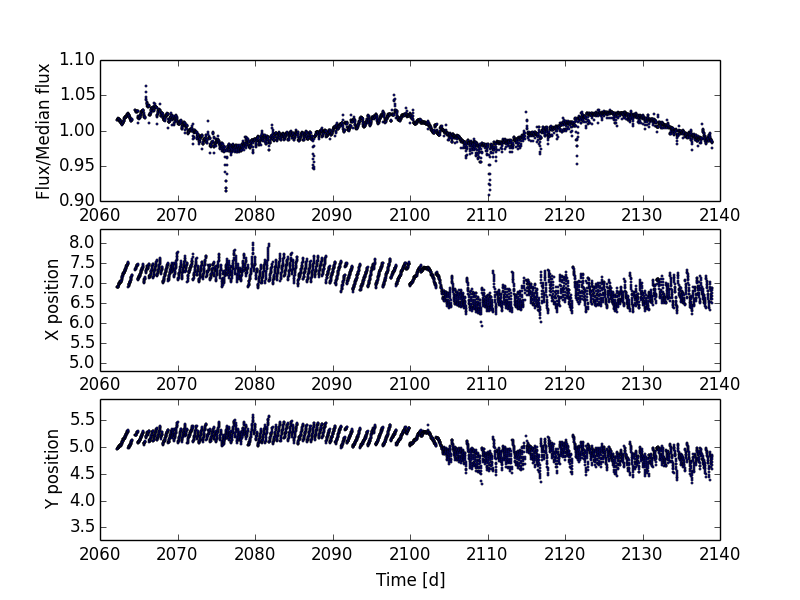}
 \caption{Aperture photometry results for the young star UScoCTIO 5. Time is specified in units of BJD-2454833. Top: Normalized light curve extracted from aperture photometry, without any subsequent detrending. The eclipse events can be seen at days 2077, 2088, 2110, and 2121. Middle and Bottom: X and Y centroid positions, in pixels, as a function of time. The six-hour interval between thruster firings (which reset the telescope position) is evident in the positions, and the position information can be used to detrend flux variations as the target moves across the detector.}\label{fig:rawcurve}
  \end{figure*}
  
   \begin{figure*}
 \epsscale{0.95}
 \plotone{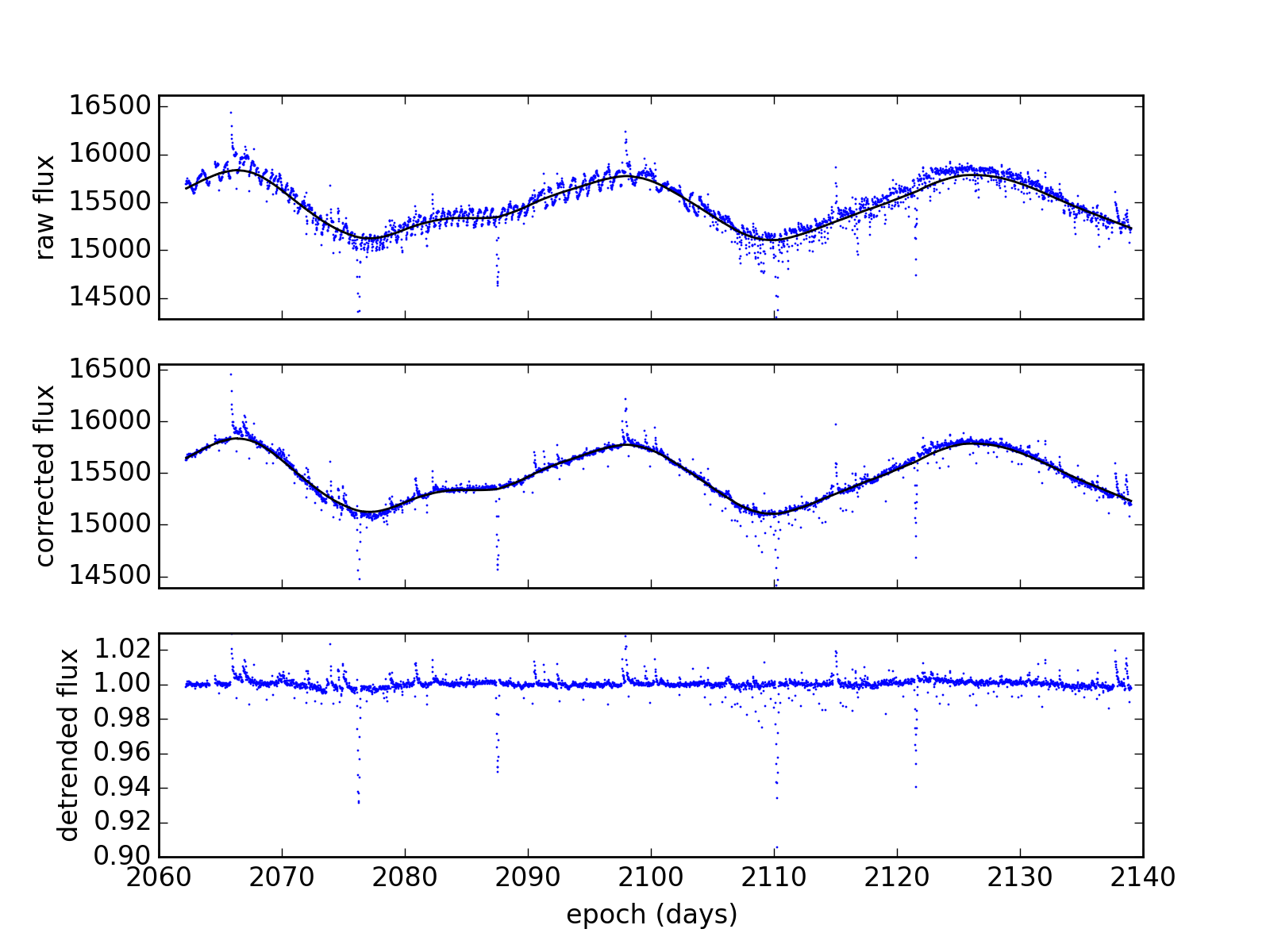}
 \caption{Light curve for UScoCTIO 5 against time in units of BJD-2454833. Top: The raw fluxes measured via aperture photometry. Middle: Corrected fluxes where the detector-position-dependent brightness changes have been detrended out. Bottom: Corrected fluxes where the out-of-eclipse variations due to spots have been detrended out. The variability is clearly dominated by spot-driven variations on a characteristic timescale of $\sim$34 days and with amplitude of $\pm$5\%, albeit with clear changes in shape between subsequent cycles. The two primary and two secondary eclipse events, which have depths of $\sim$5\% and 7\% respectively, are clearly seen. Based on detailed inspection of the light curve, the numerous brightening events appear to be astrophysical (stellar flares) and not systematic errors.} \label{fig:detrended}
  \end{figure*}
 
\subsection{K2 Photometry}

We downloaded the target pixel files (TPFs) for UScoCTIO 5 from the Mikulski Archive for Space Telescopes (MAST), where it is stored under its Ecliptic Plane Input Catalog (EPIC) identification number 205030103. This data consists of 3811 $\sim$10$\times$12 pixel stamp images centered on UScoCTIO 5, acquired between 2014 August 23 and 10 November.

Telescope pointing for K2 is unstable, due to the failure of a second reaction wheel during the Kepler prime mission in May of 2013. As a result, the centroid of UScoCTIO 5 drifts at a rate of $\sim$0.1\arcsec, or $\sim$0.02 pixels, per hour. While this movement is relatively small, associated detector sensitivity variations at the few percent per pixel level compromise the otherwise exquisite photometry. To compensate for these effects, onboard software checks the pointing every six hours and initiates a thruster firing if the roll angle has exceeded a threshold. The result is jumps in position (and hence measured flux) on this timescale. 

To minimize the effect of jumps on the photometry, we determined the stellar centroid position independently for each pixel stamp. TPF headers provide a rough world coordinate system solution which is the same for all images, but these are not precise enough to center the target. We therefore cut out a 7$\times$7 pixel substamp around the nominal target position (i.e., the green dot in the left side of Figure~\ref{fig:pics}), and determined a flux-weighted centroid from these pixels. The drift in these centroids over time can be seen in the bottom panels of Figure~\ref{fig:rawcurve}.

We then placed photometric apertures at each centroid location, extracting photometry with a set of circular apertures from 1.0 to 4.0 pixels in radius, at intervals of 0.5 pixels, and subtracting the background as determined from a wider annulus. The circular aperture is assumed to have a boundary that intersects the square pixels along an arc, with fractional flux per pixel integrated geometrically for pixels neither fully inside nor outside the circular aperture. We found that photometric noise levels after detrending for position jump effects (see Section~2.2) were minimized with the 1.5-pixel aperture. This size has the additional advantage of being small enough so as to avoid flux contamination from other stars lying $\sim$10\arcsec\, away. The chosen aperture is shown as a red circle in Figure~\ref{fig:pics}.

To remove errant data, we discarded the first 93 light curve points, for which the pixel positions were particularly different compared to the rest of the time series. We also removed points with detector anomaly flags. Finally, we pruned points lying more than five standard deviations off the median light curve trend (excluding points within or around eclipses). The resulting raw light curve is displayed in Figure~\ref{fig:rawcurve}. 

\subsection{Detrending of Instrumental Effects and Stellar Variability}

Before eclipse fitting was performed, the K2 light curve of UScoCTIO 5 was corrected for instrumental artifacts and detrended to remove starspot signatures that dominate UScoCTIO 5's out-of-eclipse light curve.  The degradation of Kepler's pointing control in the K2 mission results in light curves which, in many cases, are dominated by the target star's path across small (1-2\%) sensitivity variations in Kepler's detector. However, as \citet{Vanderburg:2014rr} and \citet{Vanderburg:2015cr} have shown, the well-behaved nature of this correlation allows the removal of much of this signature by applying a position-dependent correction to each K2 light curve. 

We applied such a correction to UScoCTIO 5's light curve, removing the systematic structure associated with the drift of the K2 focal plane.  To infer the correlation between UScoCTIO 5's xy position on the detector and the sensitivity of the K2 light curve, we first performed an initial detrending of UScoCTIO 5's light curve to remove the intrinsic stellar variations that is ubiquitous to young stars (e.g., \citealt{Cody:2010lq,Cody:2013uq,Cody:2014uq}) and whose amplitude is ~10x larger than the systematic pointing artifacts.  We removed these stellar variations using the SuperSmoother algorithm, originally developed by Friedman (1984) and subsequently implemented in python by Vanderplas (2015; 10.5281/zenodo.14475), by dividing the raw light curve by an alpha=0 supersmoothed fit. The alpha parameter provides a mechanism for biasing the SuperSmoother fit against high frequency components (i.e., providing 'bass enhancement' in Friedman's original treatment); an alpha=0 fit corresponds to an unbiased fit, which preserves the ability to fit and correct for high frequency stellar variations in the target light curve. 

With the large-scale stellar variability signatures removed, the systematic structure due to K2's pointing drift dominated the normalized light curve; we removed this structure by dividing the flux in each epoch of UScoCTIO 5's raw light curve by the median normalized flux of the 10 data points in the normalized light curve whose xy positions are closest to that of the epoch in question. The correlation between K2's sensitivity and UScoCTIO 5's xy position underwent a clear change halfway through campaign 2, when the direction of the torque resulting from the incident solar flux changed to the opposite direction. As a result, we corrected the portions of the light curve taken before and after epoch 2102 independently.\footnote{In the text and in Figures 2 and 3, we use the Kepler time coordinate system for the K2 light curve where epoch = BJD - 2454833.}

Finally, having applied this pointing correction to USco CTIO 5's raw light curve, we then again used an $\alpha=$ 0 SuperSmoother fit to detrend the light curve and remove the large-scale starspot signatures prior to fitting the eclipse profiles. The remaining light curve still showed long-term variations at the level of $\pm$3 millimagnitudes, so we removed this remaining small signal by fitting (using datapoints outside eclipse) and subtracting a DC offset  for $\pm$12 hours around each eclipse. We show the raw, instrumental-detrended, and instrumental/stellar-detrended light curves in Figure~\ref{fig:detrended}.

\section{High-Resolution Spectroscopy and Radial Velocities}

\begin{deluxetable*}{lccrrrrrrr}
\tabletypesize{\footnotesize}
\tablewidth{0pt}
\tablecaption{Keck-I/HIRES RVs}
\tablehead{
\colhead{Epoch} & \colhead{Epoch} & \colhead{Wavelength} &  \colhead{$t_{int}$)} & \colhead{SNR\tablenotemark{a}} &  \colhead{$v_{p}$} & \colhead{$\sigma_{v_p}$\tablenotemark{b}} & \colhead{$v_{s}$} & \colhead{$\sigma_{v_s}$\tablenotemark{b}} & \colhead{$F_s/F_p$}
\\
\colhead{(UT Date)} & \colhead{(HJD-2450000)} & \colhead{Range (\AA)} & \colhead{(sec)} & \colhead{} & \colhead{(km/s)} & \colhead{(km/s)} & \colhead{(km/s)} & \colhead{(km/s)} 
}
\startdata

20030611 & 2801.94316 & 6400-8700 &  900 & 49 &  -36.78$\pm$ 0.18 &    8.05$\pm$ 0.10 &   33.22$\pm$ 0.17 &    8.09$\pm$ 0.27 &  0.915$\pm$ 0.034 \\
20040509 & 3134.96680 & 7000-9250 &  900 & 50 &   -9.18$\pm$ 0.12 &    6.65$\pm$ 0.32 &    4.29$\pm$ 0.13 &    7.18$\pm$ 0.24 &  1.045$\pm$ 0.052 \\
20040510 & 3136.00935 & 7000-9250 & 1200 & 52 &  -15.49$\pm$ 0.09 &    6.92$\pm$ 0.22 &   11.47$\pm$ 0.21 &    7.01$\pm$ 0.27 &  0.941$\pm$ 0.011 \\
20040511 & 3136.90724 & 6400-8700 & 1200 & 43 &  -22.05$\pm$ 0.16 &    7.71$\pm$ 0.23 &   17.25$\pm$ 0.19 &    7.27$\pm$ 0.28 &  0.908$\pm$ 0.026 \\
20040511 & 3137.00804 & 6440-8750 &  900 & 39 &  -22.53$\pm$ 0.10 &    7.80$\pm$ 0.14 &   18.27$\pm$ 0.11 &    7.66$\pm$ 0.14 &  0.932$\pm$ 0.020 \\
20040511 & 3137.13712 & 6440-8750 &  400 &  5 &  -24.44$\pm$ 0.17 &    7.74$\pm$ 0.21 &   18.57$\pm$ 0.16 &    7.52$\pm$ 0.31 &  0.951$\pm$ 0.044 \\
20040513 & 3139.13254 & 4480-6890 &  400 & 17 &  -35.89$\pm$ 0.20 &    7.59$\pm$ 0.21 &   31.65$\pm$ 0.20 &    8.04$\pm$ 0.19 &  0.932$\pm$ 0.034 \\
20050302 & 3432.06727 & 5720-8570 &  900 & 33 &   17.79$\pm$ 0.13 &    7.85$\pm$ 0.13 &  -24.12$\pm$ 0.11 &    7.87$\pm$ 0.12 &  0.922$\pm$ 0.017 \\

\hline
Gl 83.1 & 5741.13595 & 4320-8750 & 120 & 46 & .. & .. & .. & .. & .. \\
Gl 83.1 & 5930.69346 & 4320-8750 & 120 & 71 & .. & .. & .. & .. & ..  \\
Gl 447 & 5933.16968 & 4320-8750 & 120 & 130 & .. & .. & .. & .. & ..  \\
HZ 44 & 5931.18067 & 4320-8750 & 120 & 43 & .. & .. & .. & .. & .. \\

\enddata
\tablenotetext{a}{Measurements of the spectrum's SNR and the components' flux ratio are made at $\lambda = 6600$\AA\ if that wavelength is included in the observation, and at $\lambda = 7000$\AA \ otherwise.}
\tablenotetext{b}{We report $\sigma_{v_p}$ and $\sigma_{v_s}$ as the standard deviation of the Gaussian fits to the two stars' broadening functions, which is a measure of both the instrumental broadening and the rotational broadening.}
\end{deluxetable*}

\begin{deluxetable}{lcccc}
\tabletypesize{\footnotesize}
\tablewidth{0pt}
\tablecaption{Keck-I/HIRES Spectral Features}
\tablehead{
\colhead{Epoch} & \colhead{HJD} & \colhead{$EW[H\alpha]$} & \colhead{$EW[Li]_{prim}$} & \colhead{$EW[Li]_{sec}$} 
\\
\colhead{(UT Date)} & \colhead{(days)} & \colhead{(\AA)} & \colhead{(\AA)} & \colhead{(\AA)} 
}
\startdata
20030611 & 2801.94316 & -6.1 & 0.32 & 0.23 \\
20040511 & 3136.90724 & -4.8 & 0.29 & 0.27 \\
20040511 & 3137.00804 & -3.8 & 0.31 & 0.29 \\
20040511 & 3137.13712 & -4.0 & 0.25 & 0.20 \\
20040513 & 3139.13254 & -4.7 & 0.32 & 0.27 \\
20050302 & 3432.06727 & -4.6 & 0.34 & 0.30 \\
\enddata
\end{deluxetable}

 \begin{figure*}
 \includegraphics[scale=0.90,trim=0cm 1cm 0cm 0cm]{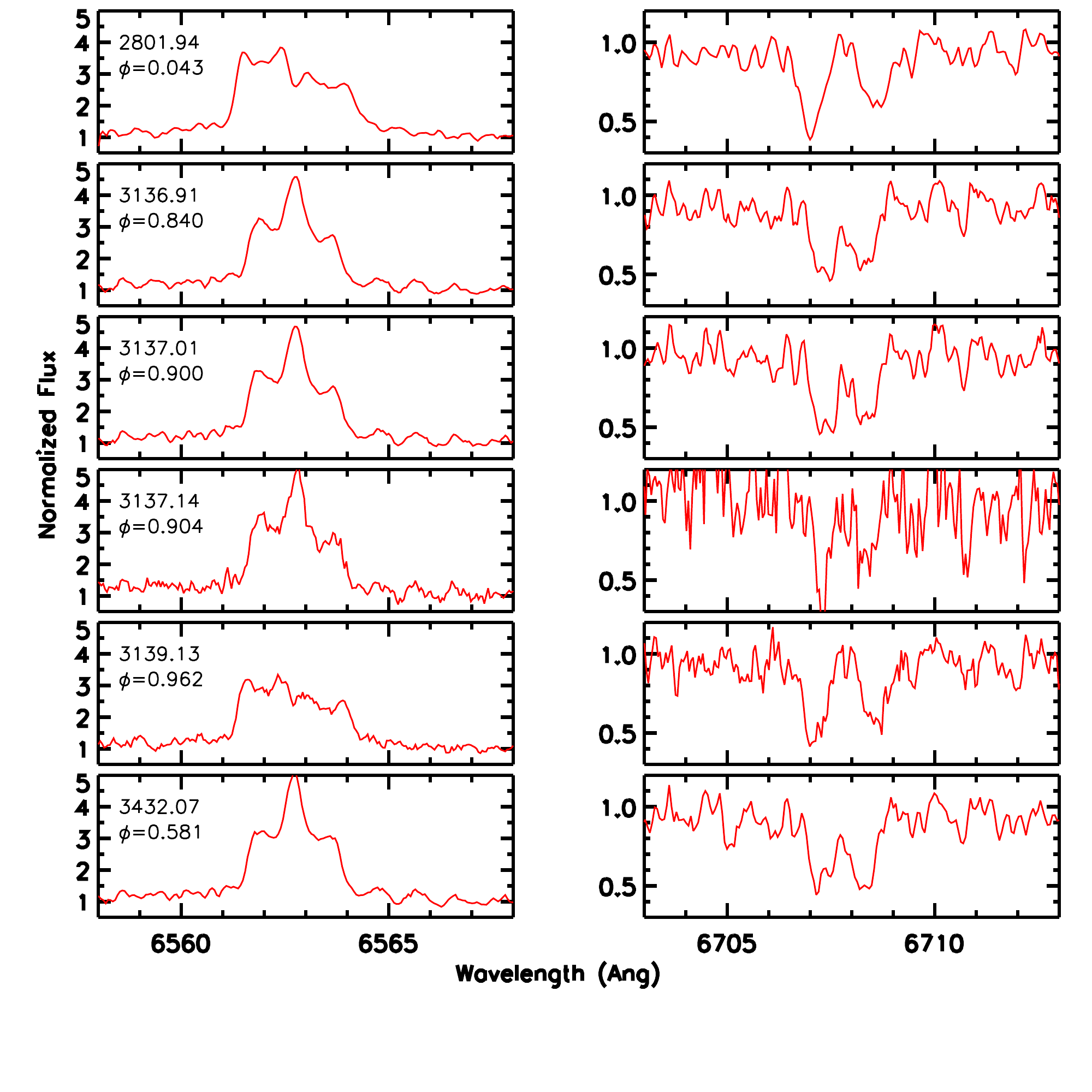}
 \caption{H$\alpha$ emission and lithium absorption lines from the six Keck/HIRES spectra that included the appropriate wavelength ranges. All spectra are normalized to the (pseudo-)continuum surrounding that line. All were observed near quadrature, resulting in clearly resolved lithium lines. However, the H$\alpha$ emission lines are never more than moderately resolved; they are broader because they are formed in the hot chromosphere, unlike the photospheric absorption lines that are only rotationally broadened.} \label{fig:HIRESlines}
  \end{figure*}

High-resolution spectroscopic measurements for UScoCTIO 5 were already reported by \citet{Reiners:2005wq}, who obtained 22 epochs in 2003--2005 with the optical echelle spectrographs Keck-I/HIRES, Magellan/MIKE, and Blanco/Echelle. They extracted the difference between the primary and secondary star RVs ($\Delta v = v_p - v_s$) for the system in each epoch, determined from a cross-correlation with the dwarf standard star Gl 406. They reported a best-fit orbit with $P = 33.992 \pm 0.006$ days, $(M_1+M_2) \sin(i)= 0.64 \pm 0.02 M_{\odot}$, $e = 0.276 \pm 0.008$, $a = 0.177 \pm 0.002$ AU, $\omega = 355.5 \pm 0.8$ deg, and $T_0 = 52799.974 \pm 0.002$ days (MJD). However, since the K2 light curve offers precise constraints on the period and a combination of the eccentricity and longitude of periastron, we have refit the orbit as part of our analysis.

We have adopted the Magellan and Blanco $\Delta v$ measurements directly from the analysis by \citet{Reiners:2005wq}. Based on the RMS scatter that they reported for their measurements around the best-fit orbit, the typical uncertainty in each measurement is $\sigma_{\Delta v} = 500$ m/s. For the Keck-I/HIRES data, we have downloaded the observations from the Keck archive and reanalyzed the spectra to measure absolute RVs for each component. These absolute velocities are needed in order to measure the mass ratio of the system, and hence the masses of the individual component stars.

Our analysis of the HIRES data is very similar to the methods described in \citet{Kraus:2011lr} and \citet{Kraus:2014rt}. We extracted and wavelength-calibrated each spectrum using the MAKEE pipeline\footnote{\url{http://spider.ipac.caltech.edu/staff/tab/makee}}, refining the wavelength solution by cross-correlating the 6800\AA\, or (preferably) 7600\AA\, telluric absorption bands against those of the spectrophotometric standard star HZ 44 \citep{Massey:1988pt}. For each spectrum of UScoCTIO 5, we then measured the broadening function (Rucinski 1999)\footnote{\url{http://www.astro.utoronto.ca/\~{}rucinski/SVDcookbook.html}} with respect to our own Keck/HIRES observations of two standard-stars over a total of three epochs: one observation of Gl 447, and two separate observations of Gl 83.1. Both standard stars have temperature and metallicity similar to UScoCTIO 5 (SpT=M4--M5 and $[M/H]\sim0.0$; \citealt{Rojas-Ayala:2012os}, Mann et al. 2015b). The broadening function is a better representation of the rotational broadening convolution than a cross-correlation, since it is less subject to ``peak pulling'' and produces a flatter continuum. We fit each broadening function with two Gaussian functions to determine the absolute primary and secondary star RVs ($v_p$ and $v_s$), the standard deviations of the lines due to rotation and instrumental resolution ($\sigma_{v_p}$ and $\sigma_{v_s}$), and the average flux ratio across all well-fit orders (which is estimated from the ratio of areas for the two peaks of the broadening function).

We list these measurements in Table 1. For those spectra that include the 6500-6700\AA\, region, we also report equivalent widths of H$\alpha$ emission (which was never fully resolved due to the intrinsic thermal broadening of the chromosphere, and hence we report as a single value) and Li$_{6708}$ absorption (which was resolved in all epochs) in Table 2. The equivalent widths are measured with respect to the continuum of the full composite spectrum, but individual stellar values can be determined from the flux ratio of the spectra (which is nearly constant across the entire wavelength range of these spectra). We also show narrow wavelength ranges around H$\alpha$ and Li$_{6708}$ in each spectrum in Figure~\ref{fig:HIRESlines}.

\section{Spectral Classification, Extinction, Temperature, and Luminosity}

 \begin{figure*}
 \epsscale{0.95}
\includegraphics[scale=0.47]{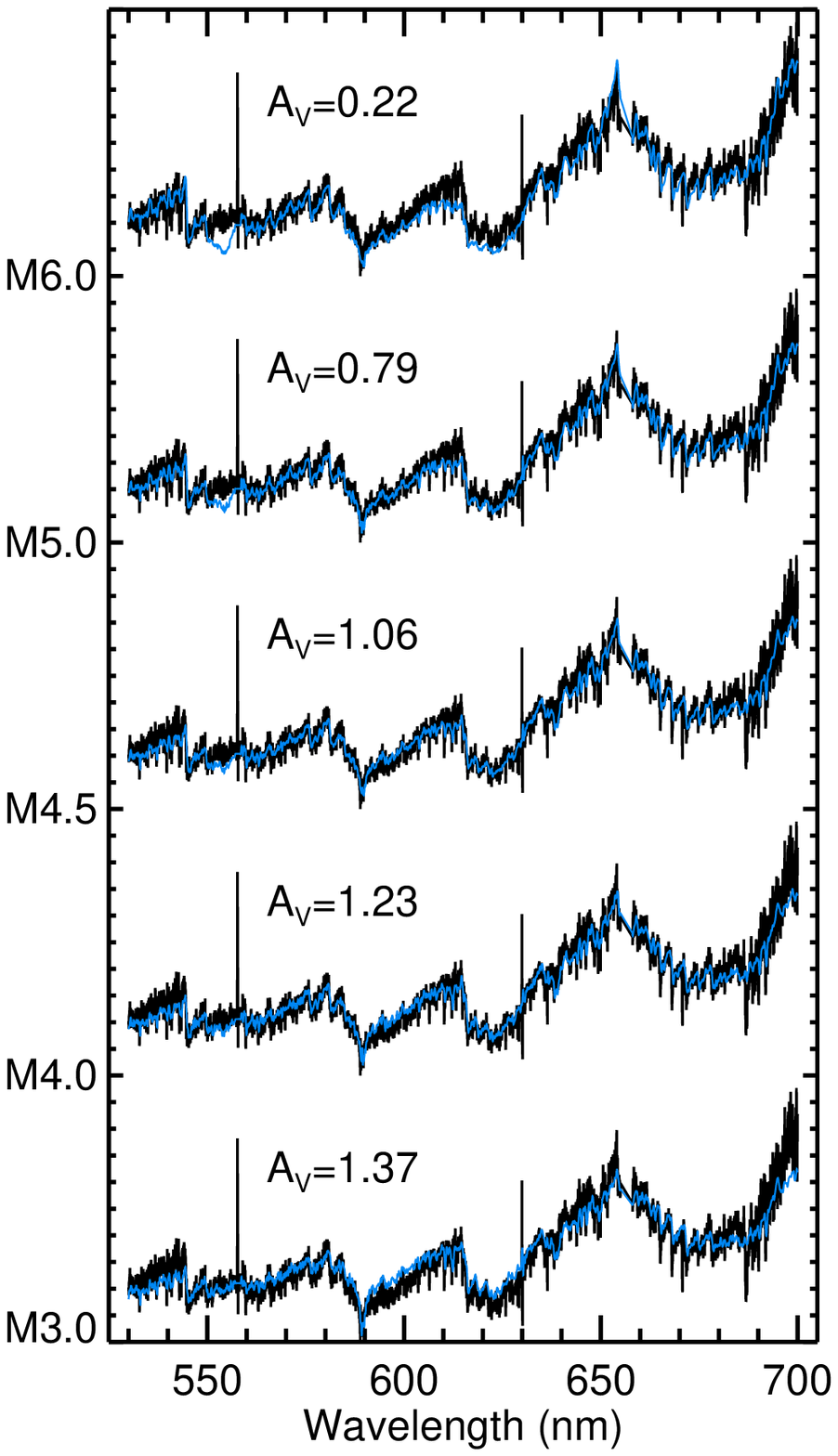}
\includegraphics[scale=0.7]{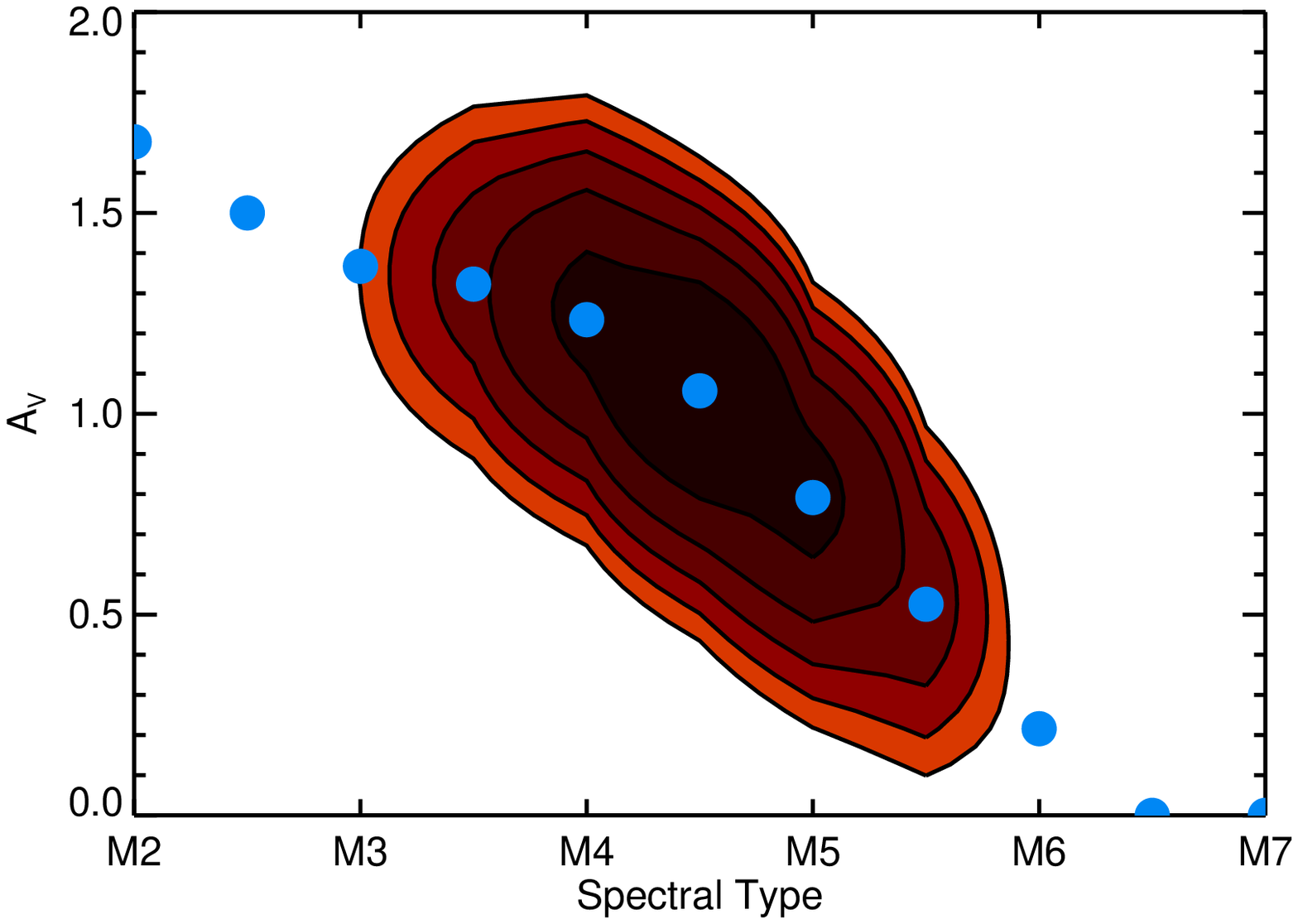}
 \caption{Left: The WIFES optical spectrum of UScoCTIO 5 as compared to a range of dwarf spectral standards. In each case, we fit for the extinction $A_V$ that minimizes the residuals; the best-fit spectrum is an M4.5V dwarf with $A_V=1.1$ added. Right: The $\chi_{\nu}^2$ surface describing an expanded range of fits like those shown. The blue dots show the best-fit value of $A_V$ for each spectral type. The contours are drawn at at levels of $\Delta \chi_{\nu} = $1, 2, 3, 4, and 5; it is difficult to infer accurate uncertainties from $\chi^2$ fits to spectra due to the strong covariances in both astrophysical and instrumental errors, but the $\Delta \chi_{\nu}^2 = 1$ contour demonstrates the range over which the models visually give a worse fit. From this and from visual inspection of the left panel, the uncertainties are $\pm$0.5 subclass in SpT and $\pm$0.3 mag in $A_V$.} \label{fig:wifes}
  \end{figure*}

\begin{deluxetable}{lrl}
\tabletypesize{\footnotesize}
\tablewidth{0pt}
\tablecaption{System Photometry}
\tablehead{
\colhead{Filter} & \colhead{$m$ (mag)} & \colhead{Reference}
}
\startdata

$B$ & 17.806 $\pm$ 0.203 & APASS \citep{Henden:2012dq}\\
$V$ & 16.192 $\pm$ 0.010 & APASS \citep{Henden:2012dq}\\
$g'$ & 16.975 $\pm$ 0.074 & APASS \citep{Henden:2012dq}\\
$r'$ & 15.482 $\pm$ 0.042 & APASS \citep{Henden:2012dq}\\
$r'$ & 15.385 $\pm$ 0.050 & CMC15 \citep{Evans:2002sp}\\
$i'$ & 13.708 $\pm$ 0.011 & APASS \citep{Henden:2012dq}\\
$J$ & 11.172 $\pm$ 0.023 & 2MASS \citep{Cutri:2003jh} \\
$H$ & 10.445 $\pm$ 0.026 & 2MASS \citep{Cutri:2003jh} \\
$K_s$ & 10.170 $\pm$ 0.021 & 2MASS \citep{Cutri:2003jh} \\
$W1$ & 10.036 $\pm$ 0.023 & ALLWISE \citep{Cutri:2013lq}\\
$W2$ & 9.838 $\pm$ 0.020 & ALLWISE \citep{Cutri:2013lq}\\
$W3$ & 9.648 $\pm$ 0.047 & ALLWISE \citep{Cutri:2013lq}\\

\enddata
\end{deluxetable}

 \begin{figure}
 \epsscale{1.12}
 \plotone{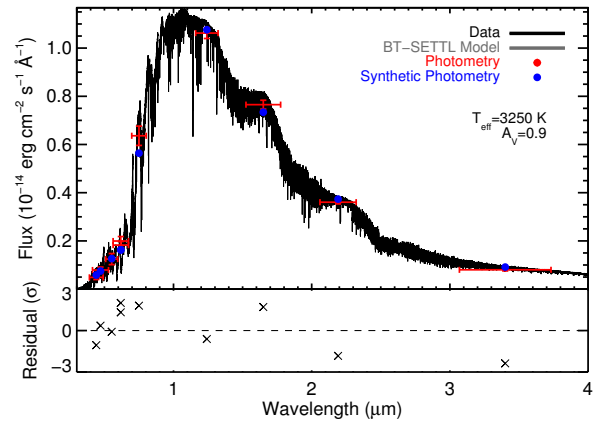}
 \caption{Photometric SED of the (unresolved) UScoCTIO 5 system (red points), as compared to the reddened best-fitting BT-SETTL model ($T=3250$ K, $A_V$=0.9) and the corresponding synthetic photometry in those bands. The bottom panel shows the difference between the synthetic and observed photometry in standard deviations.}\label{fig:fbol}
  \end{figure}

\subsection{Intermediate-Resolution Spectroscopy}
 
To better determine its spectral type and extinction, we observed UScoCTIO 5 on 2015 April 1 using the Wide-Field Spectrograph (WIFES) on the ANU 2.3m telescope. These observations are identical to those described by \citet{Rizzuto:2015fj}, which describes the data and procedures in more detail. WIFES is an integral field spectrograph with an FOV of 38$\times$25\arcsec, feeding red and blue arms. We configured the red arm to deliver a spectral resolution of $R=7000$ across a wavelength range of $\lambda=$5300--7000\AA; the flux to the blue arm was not sufficient to deliver useful data. We observed UScoCTIO 5 with an integration time of $t=500$\,s, yielding $S/N=50$ at $\lambda = $6600\AA. We processed the raw WIFES data with the ``WiFeS PyPeline''  \citep{Childress:2014lq} software in order to extract a spectral data cube. We then measured the flux in each spectral channel using PSF-fitting photometry with a Moffat profile, integrating the profile to measure the total flux from the science target in each channel and subtracting the sky background implied from the DC offset of the PSF fit. We show the extracted spectrum in Figure~\ref{fig:wifes} (left); for further detail regarding the extraction process see \citet{Rizzuto:2015fj}.

\subsection{Archival Photometry}

The geometric determination of radii offers the intriguing possibility of directly measuring empirical temperatures, as long as an accurate luminosity can be determined. While it is possible to estimate the stellar luminosity from a single flux and a corresponding bolometric correction, it should be more accurate to compile a broadband SED and add up the flux. This strategy also minimizes model-dependent uncertainties in the assumed bolometric correction. To that end, we also have compiled all of the available (component unresolved) photometry in all-sky surveys. As we summarize in Table 3, we have used photometry from 2MASS ($J$,$H$,$K_s$; \citealt{Cutri:2003jh}), AllWISE ($W1$, $W2$, $W3$; \citealt{Cutri:2013lq}), CMC15 ($r'$; \citealt{Evans:2002sp}), and APASS ($B$, $V$, $g'$, $r'$, $i'$; \citealt{Henden:2012dq}).

\subsection{Atmospheric Properties}

We have used the low-resolution optical spectrum from WIFES (Section 4.1; Figure~\ref{fig:wifes}, left) to calculate a joint constraint on the spectral type and extinction of the UScoCTIO 5 system. We compared the observed spectrum to a sequence of field M dwarfs compiled from SDSS \citep{Bochanski:2007ky}, that we artificially reddened using an $R_V=3.1$ reddening law \citep{Savage:1979jk}; the best spectrum is an M4.5 dwarf that has been reddened by $A_V=1.06$ mag. In Figure~\ref{fig:wifes} (right), we show the best-fitting $A_V$ value as a function of spectral type for the full grid of standards, as well as contours in the $\chi_{\nu}^2$ surface. It is not straightforward to convert these contours to a statistically robust confidence interval because spectra show strong covariances across wide wavelength ranges due to both astrophysical and instrumental errors. However, the interval of $\Delta \chi_{\nu}^2 = 1$ denotes the range across which the residuals become visually apparent, and hence represent an upper limit on the uncertainty in the spectral classification. Since spectral types themselves are only defined by half-subclasses, then we assess a final uncertainty of $\pm$0.5 subclass on the spectral type and a corresponding uncertainty of $^{+0.17}_{-0.27}$ mag on the extinction.

The conversion from spectral type to temperature has been a matter of longstanding debate in the star formation community. It is well known that this conversion is gravity-sensitive for M dwarfs (e.g., \citealt{Luhman:1999kl}) and perhaps even for earlier-type stars \citep{Pecaut:2013zr}, such that giants of equivalent spectral type are several hundred degrees hotter than dwarfs. L03 proposed an ``intermediate gravity'' temperature scale that might be appropriate for young stars, constructed from an interpolation of the dwarf and giant scales that makes the GG Tau system look coeval (as per \citealt{White:1999en}). Given the spectral type of M4.5$\pm$0.5, the inferred temperature on this intermediate-gravity scale is $T_{\rm eff} = 3200 \pm 75$ K. The dwarf scales would imply a lower temperature of $T_{\rm eff} = 3100 \pm 90$ K according to \citet{Leggett:1996zf}, or warmer temperatures of $T_{\rm eff} = 3190 \pm 75$ K from more recent scales (\citealt{Rajpurohit:2013qy}; Mann et al. 2015b) while the giant scale \citep{van-Belle:1999yc} would imply a higher temperature of $T_{\rm eff} = 3435 \pm 55$ K. \citet{Herczeg:2014oq} more fully review the wide range of other proposed temperature scales, which yield temperature estimates spanning 3080--3305 K for an M4.5 dwarf star; they develop an independent young star temperature scale that would predict $T_{\rm eff} = 3085$ K for UScoCTIO 5. For this work, we adopt the temperature scale of L03 and the corresponding spectroscopic temperature of $T_{\rm eff} = 3200 \pm 75$ K in order to remain consistent with the previous literature, but we also remind the reader that substantial systematic uncertainties remain.

As we discuss further in Section 6.2, it is possible to empirically anchor the SpT-$T_{\rm eff}$ relation at M4.5 using our geometric radius measurements if we can also measure the luminosity. To that end, we computed $F_{bol}$ following the procedure from Mann et al. (2015b). We used a set of solar-metallicity, $\log(g)=4.25$ BT-SETTL models \citep{Allard:2011qf} built with the \citet{Caffau:2011qy} solar abundances, which better reproduce the observed spectra of field M dwarfs than BT-SETTL models utilizing the \citet{Asplund:2009lr} abundances \citep{Mann:2013fj}. We selected grid points spanning the range of temperatures found above, each of which we reddened by a grid of extinction values consistent with the value and error derived using the empirical spectrum. We then scaled each model spectrum to give the best agreement (lowest $\chi^2$) between synthetic and observed photometry and integrate over the (un-reddened but scaled) model spectra to determine $F_{bol}$. The range of temperatures and extinction values produced a range of $F_{bol}$ values, which we use as an estimate of the uncertainty. The best fit model ($T_{model}=3250$ K, $A_V=0.9$ mag) is shown in Figure~\ref{fig:fbol}. To convert from $F_{bol}$ to $L_{bol}$, we assume a distance of $d = 145 \pm 15$ pc, consistent with the mean and scatter seen for BAF stars in Upper Scorpius (e.g., \citealt{de-Zeeuw:1999xv, Rizzuto:2011cr}). The $\sim$20\% uncertainty in extinction dominates the uncertainty in $F_{bol}$ ($\sigma_{Fbol} \sim$5\%). The 10\% uncertainty in the distance, which propagates to a 20\% uncertainty in the luminosity, represents the dominant source of uncertainty in $L_{bol}$. Using the model to fill the gaps between the photometry, we find a total bolometric flux of $F_{bol}=(2.02^{+0.13}_{-0.08}) \times 10^{-10}$ erg/s/cm$^2$. For an assumed distance of $d = 145 \pm 15$ pc, the corresponding luminosity is $L_{tot} = 0.132^{+0.028}_{-0.030} L_{\odot}$, or individual component luminosities of $L \sim 0.065 L_{\odot}$.

\section{Results}

\begin{deluxetable}{lr}
\tabletypesize{\footnotesize}
\tablewidth{0pt}
\tablecaption{System Parameters for UScoCTIO 5}
\startdata
\hline \multicolumn{2}{l}{\it Orbital Parameters}\\ \hline
                          $T_0$ (HJD) &       56914.490 $\pm$          0.026 \\
                           $P$ (days) &        34.00073 $\pm$        0.00007 \\
                             $a$ (AU) &         0.17749 $\pm$        0.00031 \\
                                  $e$ &         0.26741 $\pm$        0.00011 \\
                            $i$ (deg) &          87.912 $\pm$          0.010 \\
                       $\omega$ (deg) &          355.13 $\pm$           0.30 \\
                      $\gamma$ (km/s) &           -2.64 $\pm$           0.07 \\
\hline \multicolumn{2}{l}{Stellar Bulk Parameters}\\ \hline
              $M_p+M_s$ ($M_{\odot}$) &          0.6452 $\pm$         0.0034 \\
                        $q = M_s/M_p$ &           0.963 $\pm$          0.007 \\
                  $M_p$ ($M_{\odot}$) &          0.3287 $\pm$         0.0024 \\
                  $M_s$ ($M_{\odot}$) &          0.3165 $\pm$         0.0016 \\
              $R_p+R_s$ ($R_{\odot}$) &           1.644 $\pm$          0.008 \\
                            $R_s/R_p$ &           0.972 $\pm$          0.011 \\
                  $R_p$ ($R_{\odot}$) &           0.834 $\pm$          0.006 \\
                  $R_s$ ($R_{\odot}$) &           0.810 $\pm$          0.006 \\
	         $\log(g_p)$ (cm/s$^2$) &            4.11 $\pm$           0.01 \\
                 $\log(g_s)$ (cm/s$^2$) &            4.12 $\pm$           0.01 \\
\hline \multicolumn{2}{l}{Stellar Atmospheric Parameters}\\ \hline
                            $F_s/F_p$ &           0.999 $\pm$          0.017 \\
                            $T_s/T_p$ &           1.000 $\pm$          0.004 \\

                          $A_V$ (mag) &                  1.06$^{+0.17}_{-0.27}$ \\
                                  SpT &            M4.5 $\pm$            0.5 \\
                    $T_{\rm eff,L03}$ (K) &            3200 $\pm$             75 \\
                    $T_{\rm eff,HH14}$ (K) &            3085 $\pm$             105 \\
                   $F_{bol}$ (erg/s/cm$^2$) &    $(2.02 ^{+0.13}_{-0.08}) \times 10^{-10}$ \\
\hline \multicolumn{2}{l}{Parameters Using Distance}\\ \hline
$L_{bol}$ ($L_{\odot}$) &           $0.132^{+0.028}_{-0.030}$ \\
$L_{bol}$ ($L_{\odot}$) &           $0.132^{+0.009}_{-0.014} (D/145 pc)^2$ \\

                   $T_{\rm eff,geom}$ (K) &            $3235^{+160}_{-200}$ \\
                   $T_{\rm eff,geom}$ (K) &            $3235^{+50}_{-33} (D/145 pc)^2$ \\
\hline \multicolumn{2}{l}{Infrared Flux Method Distance}\\ \hline
$d_{\rm L03}$    (pc) & 144.4 $\pm$ 6.6  \\
$d_{\rm HH14}$ (pc) & 135.1 $\pm$ 8.8 \\
\enddata
\end{deluxetable}

 \begin{figure*}
 \epsscale{1.12}
 \plotone{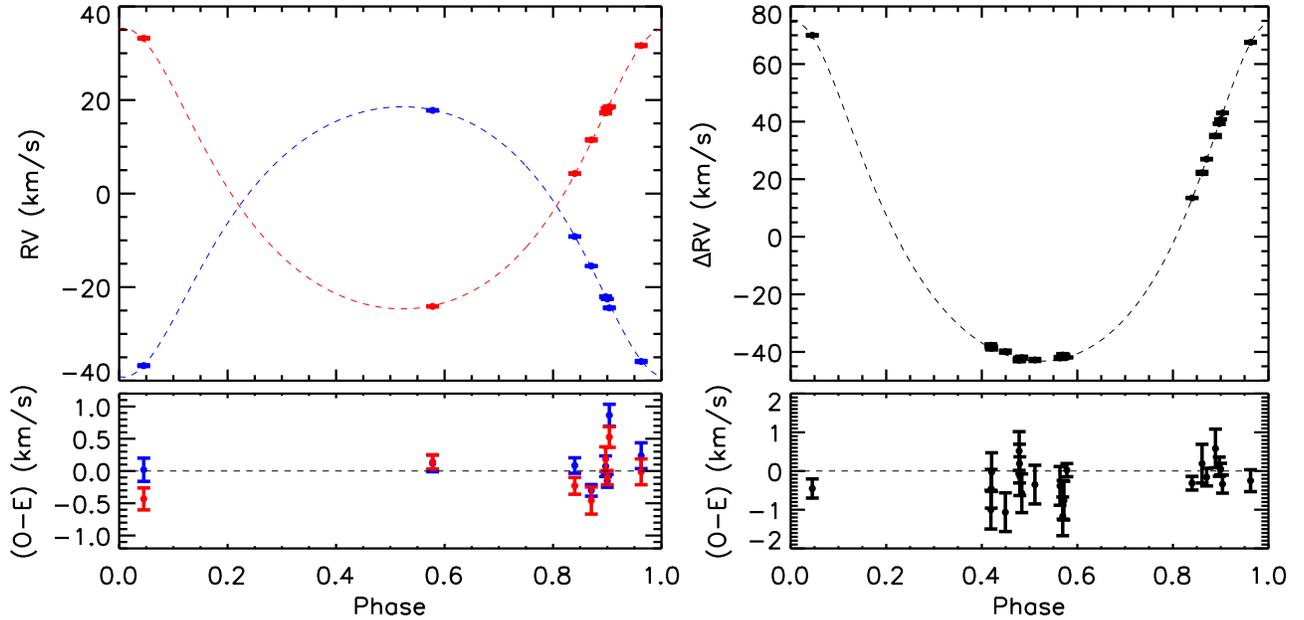}
 \caption{Left: Absolute radial velocities $v_p$ and $v_s$ for the primary and secondary stars of UScoCTIO 5, as measured from the Keck/HIRES epochs listed in Table 1. We also show the best-fit model as determined from our fitting procedure (Section 5.1). Underneath, we show the (O-E) residuals with respect to the best-fit model. Right: Relative velocity differences $\Delta v = v_s-v_p$ for all epochs, including the measurements reported by \citep{Reiners:2005wq}, also with the best-fit model curve and (O-E) residuals shown.} \label{fig:rvfit}
  \end{figure*}
  
   \begin{figure*}
 \epsscale{1.12}
 \plotone{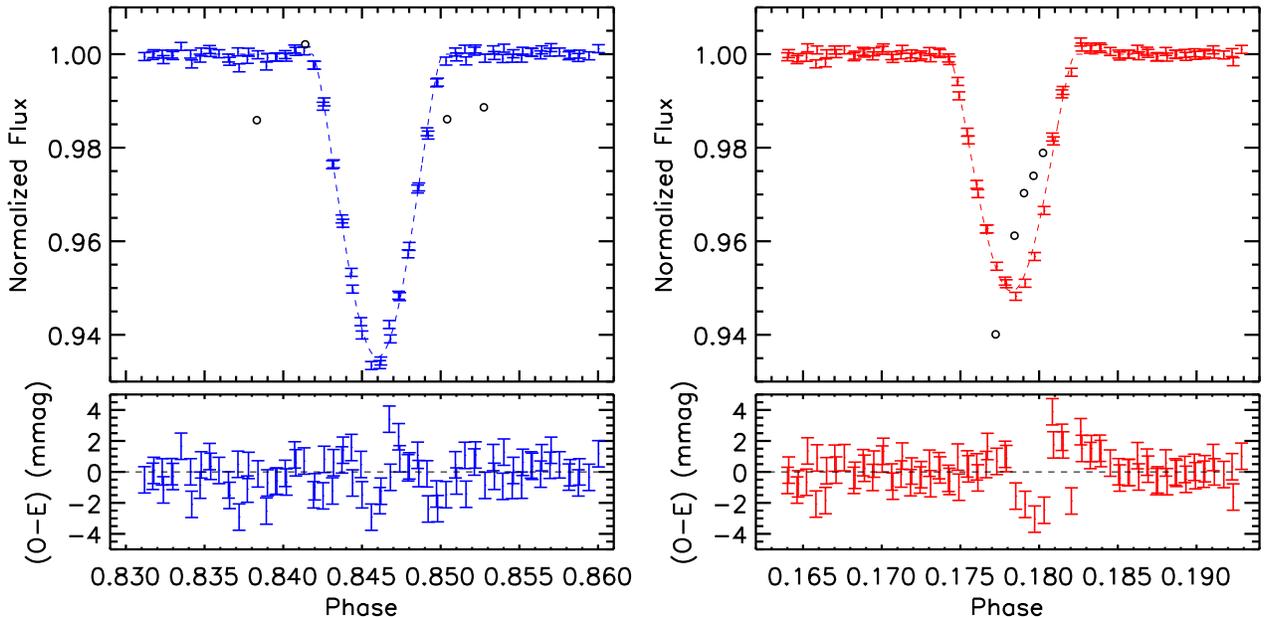}
 \caption{Primary eclipse (left) and secondary eclipse (right) for UScoCTIO 5, along with the best-fitting models (dashed lines) and the (O-E) residuals (bottom panels). Data points that were rejected as outliers are shown with open circles; a flare occurred during one secondary eclipse, so we have rejected all affected points. The primary eclipse is deeper primary due to the orbital geometry, such that a larger total surface area is occulted when the primary star is behind the secondary star; the surface brightnesses of the two stars are nearly identical. Despite the grazing nature of the eclipses, the full radii can be determined with small uncertainties and minimal covariance due to constraints from the timing of primary versus secondary eclipse, the relative eclipse durations, the relative eclipse depths, and the precise RV curve.} \label{fig:lcfit}
  \end{figure*}

\subsection{Fitting for Orbital and Stellar Parameters}

In the simple case of a tidally circularized orbit, fitting for the parameters of an eclipsing binary is a straightforward and separable problem: the component masses (moduli $\sin(i)$) can be derived from the radial velocity curve, while the inclination, period, radii, and stellar flux ratio can be derived from the light curve. However, eccentric systems are more complicated, with many covariances between parameters. For example, while the RV curve provides a direct measurement on the eccentricity $e$ and the argument of periapse $\omega$, the orbital phases of primary and secondary eclipse can provide a very tight joint constraint on both parameters. The durations of the eclipses provide a further constraint on $e$ and $\omega$, as well as posing a joint constraint on the inclination and surface brightness ratio (since the surface areas occulted can differ between the primary and secondary eclipses). Finally, since the RVs and photometry were measured nearly a decade apart, then timekeeping errors could lead to covariances in the parameters. For a system this wide, apsidal motion should be negligible on decade timescales (e.g., \citealt{Feiden:2013rm}), so we do not expect the orbital parameters to have changed between the epochs of the RV data and K2 data.

To properly account for these covariances, we have constructed a Markov-Chain Monte Carlo (MCMC) procedure to simultaneously fit both the RV curve and the light curve using a full model of all parameters. We specifically consider a model where the explicitly fit parameters are 6 orbital elements ($T_0$, $P$, $a$, $e$, $\omega$, and $i$, neglecting $\Omega$ in the absence of spatially resolved information), the mass ratio of the system $q = M_s/M_p$, the systemic radial velocity $\gamma$, the sum of the stellar radii $R_{tot}=R_p+R_s$, the ratio of the stellar radii $r=R_s/R_p$, and the ratio of stellar fluxes through the Kepler $K_p$ bandpass $f=F_s/F_p$.

We note that the convention for eclipsing binaries is not to fit for $f=F_s/F_p$, but rather for a temperature ratio or temperature difference (indeed, as we have done in the past; \citealt{Kraus:2011lr}). In principle, the ratio of surface brightnesses can be described by the ratio of temperatures; in the limiting case of a pure blackbody, the relation is exact and analytic. However, cool stellar atmospheres are notably non-blackbody, making this relation more complex. Traditional fitting codes like the Wilson-Devinney algorithm (\citealt{Wilson:1971ig}, and extensive updates thereof) parametrize this relation using Kurucz model atmospheres \citep{Kurucz:1979kl}. However, those codes typically only extend to $T \sim 3500$ K (hotter than our observed targets), and the conversion from temperature ratio to flux ratio is not easily quantified or changed. We find it more straightforward to fit directly for the flux ratio in the Kepler bandpass, and then deal with the conversion to a temperature ratio explicitly. This choice can also be found in other fitting codes such as JKTEBOP \citep{Southworth:2004ja}, which shares many design choices with our procedures.

In executing our MCMC procedure, we use analytic equations to construct a predicted RV curve against which we can compare the observations and measure the residuals. In order to isolate potentially correlated uncertainties between the measurements of the primary and secondary star at each epoch (such as from uncertainties in the wavelength scale), we chose to fit the primary star RV $v_p$ (for the 8 epochs of HIRES data) and the difference in RV $\Delta v = v_s-v_p$ (for all 22 epochs). Some fit parameters (such as the semimajor axis, and hence the total system mass) only depend on $\Delta v$, while others (such as the system velocity and the mass ratio) necessarily depend on the individual component RVs, so this choice ensures that the correlated RV errors between primary and secondary star will not unnecessarily inflate the uncertainty in parameters that do not require component-resolved measurements.

The analysis of the light curve is less straightforward. As we mention above, the Wilson-Devinney code is the gold standard of the field due to the wide range of physical effects (such as reflected light and tidal distortion) that it can reproduce. However, most of these physical effects are not needed for a binary with a semimajor axis of nearly 0.2 AU, and the architecture needed to encompass them results in a long runtime (of order seconds) to produce a single model. Furthermore, the Wilson-Devinney solution for stars that fall outside their modeled temperature range (interpolation between the coolest model and a blackbody) is also not appropriate for cool stars, meaning that its implicit conversion from $f$ to a temperature ratio would not be correct.

We therefore have instead constructed an analytic formalism that uses the work of \citet{Mandel:2002ai} to calculate the total light removed from the system due to occultation of whichever star is more distant. This analytic model can be calculated $\sim$100 times more quickly, allowing for more and longer MCMC chains that better explore the complex multi-dimensional parameter space of the fit. To account for the long duration of individual K2 exposures (which can result in a significant deviation between the midpoint flux and the average flux), we calculated the occulted flux at one minute intervals within each integration, and then calculated the average value for comparison to the observations. We chose our fit parameters to encompass known covariances (for example, fitting $R_{tot}=R_s+R_p$ and $r=R_s/R_p$, since the two radii are known to be degenerate and anti-correlated in EB fits). To allow for limb darkening, we use a quadratic relation with the coefficients prescribed for a star of appropriate $T_{\rm eff}$ and $\log(g)$ by \citet{Claret:2012tg}: $\gamma_1 = 0.5125$ and $\gamma_2 = 0.2533$.

Finally, our fit also includes the observed optical flux ratio from the Keck/HIRES spectra ($0.943 \pm 0.015$), as determined from the ratio of the integrated areas under each star's broadening function. Since the HIRES spectra represent the same wavelength range as the K2 light curve, then this comparison directly offers a joint constraint on the ratio of surface brightnesses ($f=F_s/F_p$) and the ratio of radii ($r=R_s/R_p$). As we mention above, the latter parameter in particular can be highly degenerate for grazing-eclipse systems without an extra constraint.

We executed the MCMC using a Metropolis-Hastings sampler to walk through parameter space, using jump sizes drawn from Gaussian distributions with standard deviation corresponding to a characteristic jump size. We executed several test chains early in this process, tweaking the jump sizes to yield acceptance rates of $20 < P < 50$\%. We then computed four simultaneous chains for a total length of $10^6$ steps per chain. As a result, our distributions have $4 \times 10^6$ distinct samples from which the posteriors on each parameter are constructed. We also verified that the individual chains yield values that agree to within much less than the reported $1\sigma$ uncertainties, indicating that they are well-mixed.

Finally, we calculated other parameters of interest ($M_p$, $M_s$, $M_{tot}=M_p+M_s$, $R_p$, $R_s$, $T_{\rm eff,s}/T_{\rm eff,p}$) from the fit parameters at each step in the chain, yielding similar posterior distributions. This method naturally propagated the uncertainties and covariances in the fit parameters through to the uncertainties in the derived parameters. In particular, the ratio of effective temperatures was calculated simply assuming a blackbody, since the surface brightness ratio is statistically consistent with unity and hence the stellar photospheres (and resulting emergent spectra) must be very nearly identical.

The \citet{Mandel:2002ai} formalism faces limitations for some systems, since it can't encompass reflected light, tidal distortion, or various other effects. However, most of these effects should not be relevant for a wide system like UScoCTIO 5. The one exception is the influence of spots; the 5--10\% variations in out-of-eclipse flux suggest that both stars are likely to be covered with large and complicated spot patterns. If spots are occulted during an eclipse, then the different surface brightness of the covered area will lead to a different change in the total system brightness, distorting the eclipse morphology. Indeed, we appear to see these effects in our own light curve fits at the $\pm$2 mmag level. 

The traditional solution has been to fit with a spot model (typically consisting of a few large spots) and optimize their size, latitude, longitude, and temperature to match the out-of-eclipse brightness. However, these spot models are highly degenerate, with many possible configurations replicating the same broad variations. If the incorrect spot model is used (as is almost certainly the case), then it will degrade the precision of the eclipse fit by simultaneously not encompassing the fine details of the spot structure (which can not be fit from the variations in total system flux) and forcing the fit to account for a spot model that is not correct. We therefore argue that the most conservative solution is to fit with no spots, and then forward-model a range of spot models into the observational space and determine the resulting scatter in best-fit solutions. As we discussed in \citet{Kraus:2011lr}, using this procedure for field M-dwarf eclipsing binaries with similar variations resulted in radius uncertainties of $\pm$2\%, which we adopt as a systematic uncertainty on our radii in this paper.

\subsection{System Properties} 

UScoCTIO 5 is one of the few young low-mass binaries ($\tau \la 10$ Myr; $0.1 < M_* < 0.7 M_{\odot}$) for which precise masses and radii have been determined, and therefore it represents a strong test of pre-main-sequence stellar evolutionary models. We summarize our best-fit properties of UScoCTIO 5 and its component stars in Table 4, and in Figures~\ref{fig:rvfit} and \ref{fig:lcfit}, we show the observed RVs and photometry, the best-fit model RV curve and light curve, and the residuals between the observations and the data.  We find that UScoCTIO 5 consists of two nearly identical components with masses $M \sim 0.32 M_{\odot}$ and radii $R \sim 0.82 R_{\odot}$; the system mass that we calculate is consistent with the value of $M \sin(i)$ measured by \citet{Reiners:2005wq}. The fractional uncertainties on the individual masses and radii we measure are $<$1\%, due to the precise RVs that can be obtained from Keck/HIRES and the exquisite photometry from K2. Given that the mass ratio $q$, radius ratio $r$, and surface brightness ratio $f$ are all nearly (but just under) unity, then the two components appear to be very nearly coeval.

As we summarize in Table 1, both stars have a total broadening of their spectral lines (from rotation and instrumental resolution) of $\sigma_{v} = 7.5 \pm 0.2$ km/s. The observations were taken in modes that produce a spectral resolution of $R = 36000$ (FWHM) or $\sigma_{v,inst} \sim 3.5$ km/s, implying that the stars have an intrinsic rotation of $\sigma_{v,rot} \sim 6.6$ km/s or a rotational period (given the measured radii, and an assumption of spin alignment with the orbital plane) of $P_{rot} \sim 6$ days. As can be seen in Figure~\ref{fig:detrended}, out-of-eclipse variability seems to occur on a much longer timescale. This suggests that the variability might be a result of long-term secular changes in spot coverage or a beating pattern between the two stars' complicated spot maps, rather than rotational modulation from a coherent and static spot pattern. With only two intervals of this longer-term variation available in the K2 dataset, the nature of this variability remains ambiguous.

Finally, there are now several lines of evidence demonstrating that UScoCTIO 5 is a young member of the Upper Scorpius OB association. As we show in Table 2 and \ref{fig:HIRESlines}, each star has a lithium equivalent width of $\sim$250--300 m\AA; accounting for the presence of two stars' worth of continuum flux, these values are consistent with the 500--600 m\AA\, equivalent widths presented for single M3 stars in Upper Sco (e.g. \citealt{Preibisch:2001gn,Preibisch:2002qt,Rizzuto:2015fj}). Both stars also show H$\alpha$ emission at levels that are consistent with the SpT-EW[H$\alpha$] sequence observed by \citet{Kraus:2014rt} for non-accreting young stars in the Tuc-Hor moving group, as well as frequent flaring (Figure~\ref{fig:detrended}), including one small flare during a secondary eclipse. 

Our measurement of the system radial velocity ($\gamma = -2.64 \pm 0.07$ km/s), when combined with the proper motion from UCAC4 ($\mu = (-11.0,-18.7) \pm 2.1$ mas/yr; \citealt{Zacharias:2012lr}), allows us to further test the membership using kinematics. If we assume a distance of $d = 145 \pm 15$ pc, then the corresponding space velocity with respect to the Sun is $v_{UVW} = ( +2.9\pm 0.7 , -14.2 \pm 2.0 , -4.3 \pm 1.4 )$ km/s\footnote{All measurements of $UVW$ velocities are presented in the sense that $U$ is positive in the galactic anti-center direction, as encoded in the IDL routine gal\_uvw.pro. }. The ten nearest high-confidence BAF members of Upper Sco have an average velocity of $V_{UVW} = (+1.7 \pm 0.9, -16.3 \pm 0.9, -6.8 \pm 0.6)$\,km/s; \citep{Rizzuto:2011cr}, while all of Upper Sco has an average velocity of $V_{UVW} = (+6.4 \pm 0.5, -15.9 \pm 0.7, -7.4 \pm 0.2)$\,km/s \citep{Chen:2011ys}; the difference between these measurements might point to internal kinematic substructure within Upper Sco. Our measurements are consistent with the adjacent BAF stars to within $< 2$ km/s and with the average for all of Upper Sco to within $< 4$ km/s. Given the expected internal velocity dispersion of at least 1--2 km/s on small scales and potentially more on association-wide scales (e.g., \citealt{Kraus:2008fr}), the kinematics are therefore consistent with those expected of an Upper Sco member. The proper motion alone is also consistent with the mean value of Upper Sco ($\mu = (-9.3,-20.2)\pm 0.5$ mas/yr; \citealt{Kraus:2007ve}) to within the uncertainties. We therefore further confirm that UScoCTIO 5 is both young and comoving with Upper Scorpius.

\section{Discussion}

\subsection{Comparison to Stellar Evolutionary Models}

   \begin{figure*}
 \epsscale{1.12}
 \plotone{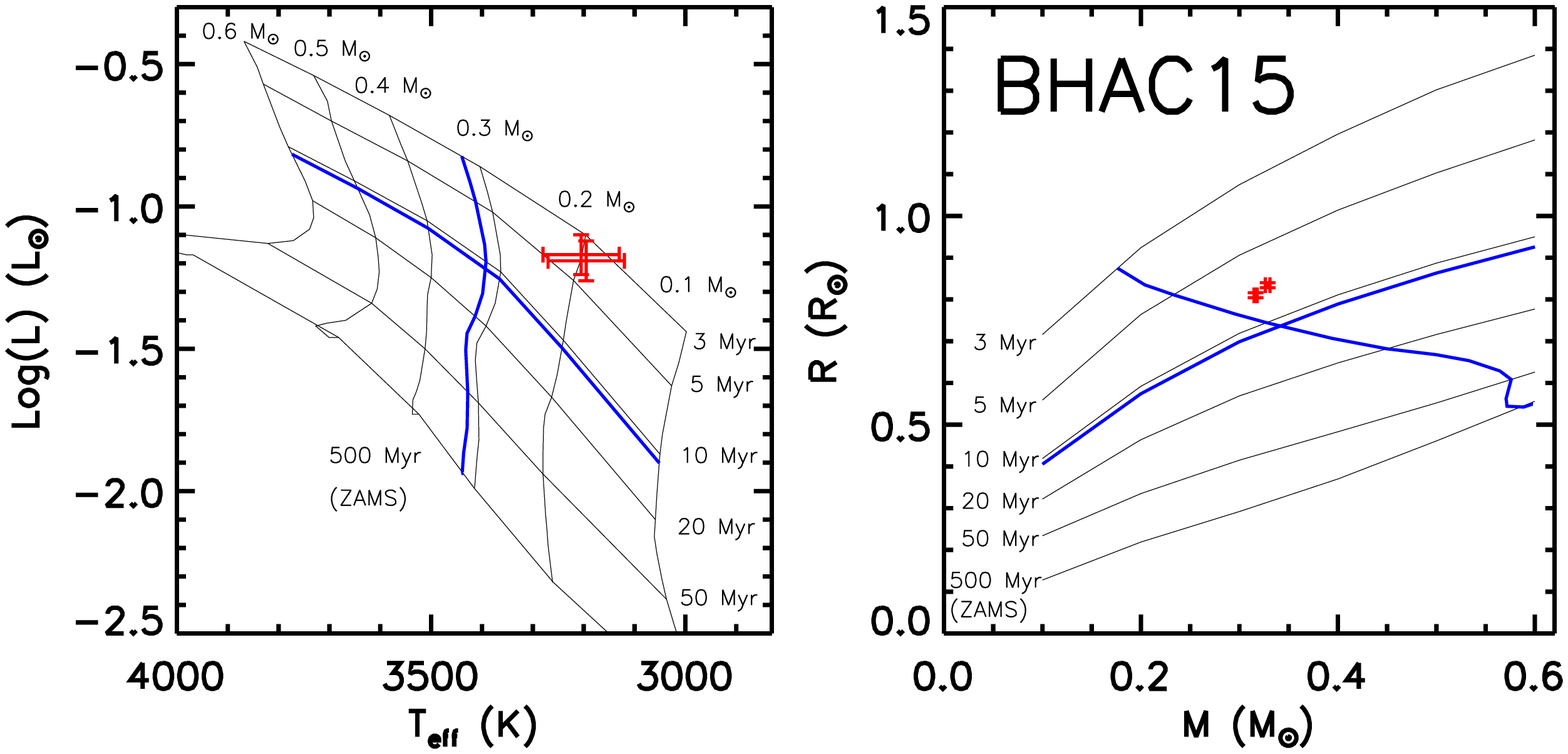}
 \caption{Left: $L$-$T_{\rm eff}$ HR diagram showing the measured positions of UScoCTIO 5 A+B (red error bars) and the isochronal and isomass sequences predicted for low-mass stars by the BHAC15 models (Baraffe et al. 2015). The components are offset slightly in $T_{\rm eff}$ for clarity. The isomass model track corresponding to the component masses ($M \sim 0.32 M_{\odot}$) and the isochrone model track corresponding to the currently accepted value for Upper Sco ($\tau \sim 11$ Myr; \citealt{Pecaut:2012dp}) are shown in blue. Perfect agreement with the models should show the components of UScoCTIO 5 sitting at the intersection of the blue lines; we find that the isomass line does not match with observations, indicating that the $T_{\rm eff}$ predicted by the models is too high. Right: Mass-Radius diagram showing the measured positions of UScoCTIO 5 A+B (red error bars) and the isochronal sequences of the BHAC15 models. As in the HR diagram, we use blue lines to show the model tracks for the expected isochrone ($\tau \sim 11$ Myr) and the luminosity that we measure ($L_{bol} \sim 0.066 L_{\odot}$ for each star). We find that the position predicted by the models (at the intersection of the blue sequences) matches the mass, but not the radius; the models predict radii that are too small, equivalent to predicting $T_{\rm eff}$ to be too high (but avoiding the systematic uncertainties of a direct comparison using $T_{\rm eff}$).} \label{fig:bhac15}
  \end{figure*}
  
  \begin{figure*}
 \epsscale{1.12}
 \plotone{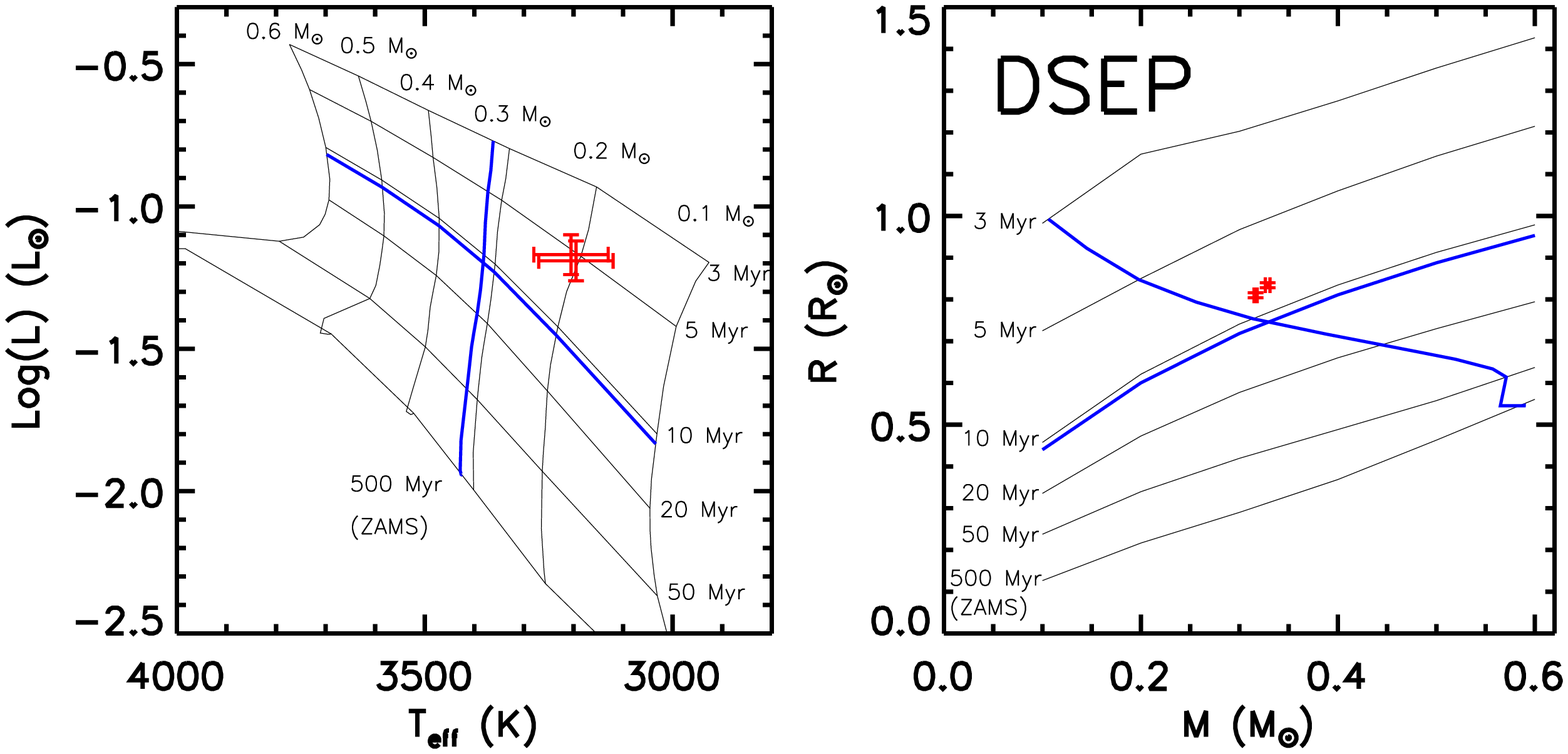}
 \caption{As in Figure~\ref{fig:bhac15}, but for the DSEP models \citep{Dotter:2008qq,Feiden:2015jk}.} \label{fig:dsep}
  \end{figure*}
  
     \begin{figure*}
 \epsscale{1.12}
 \plotone{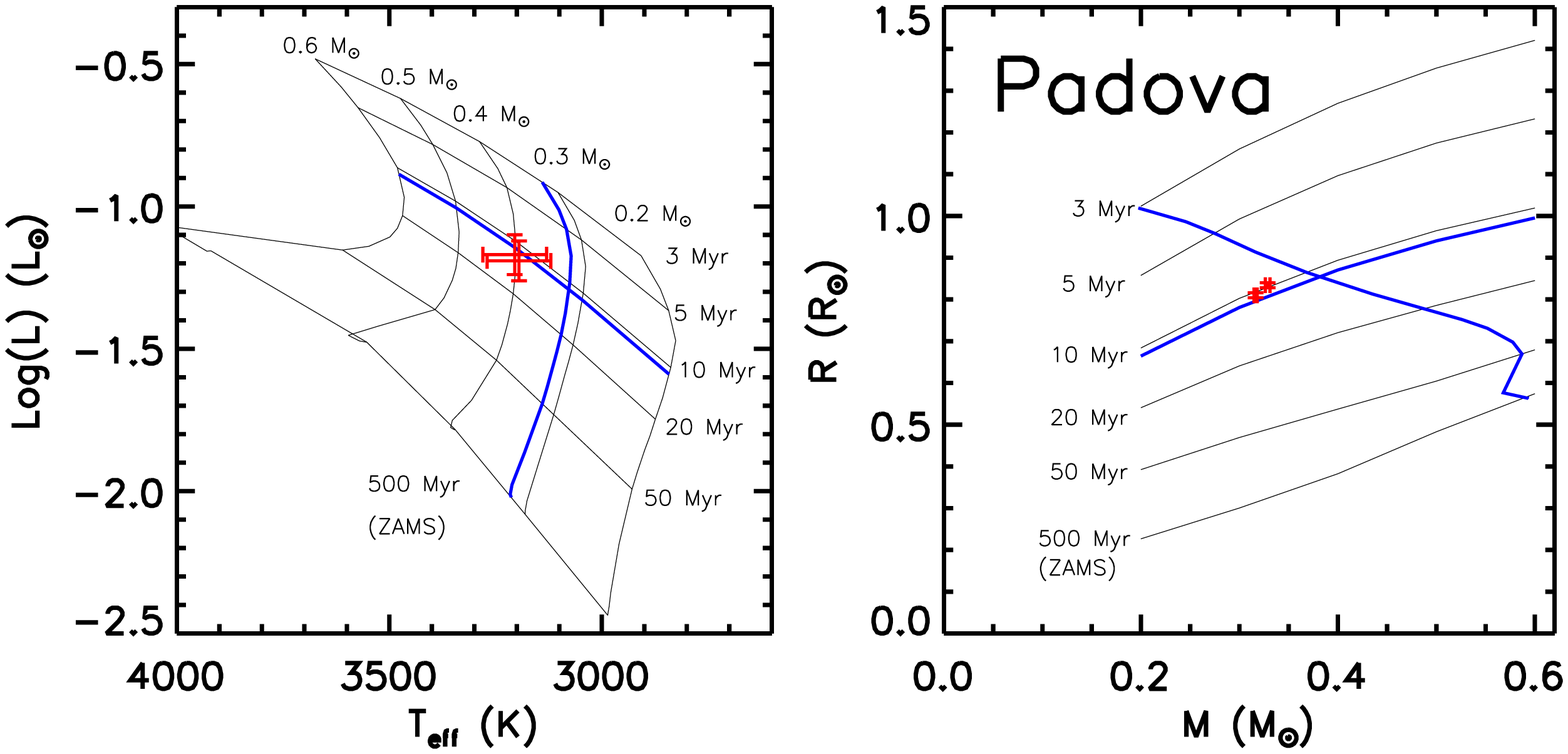}
 \caption{As in Figure~\ref{fig:bhac15}, but for the Padova/PARSEC models \citep{Bressan:2012hc,Chen:2014fk}.} \label{fig:padova}
  \end{figure*}
  
     \begin{figure*}
 \epsscale{1.12}
 \plotone{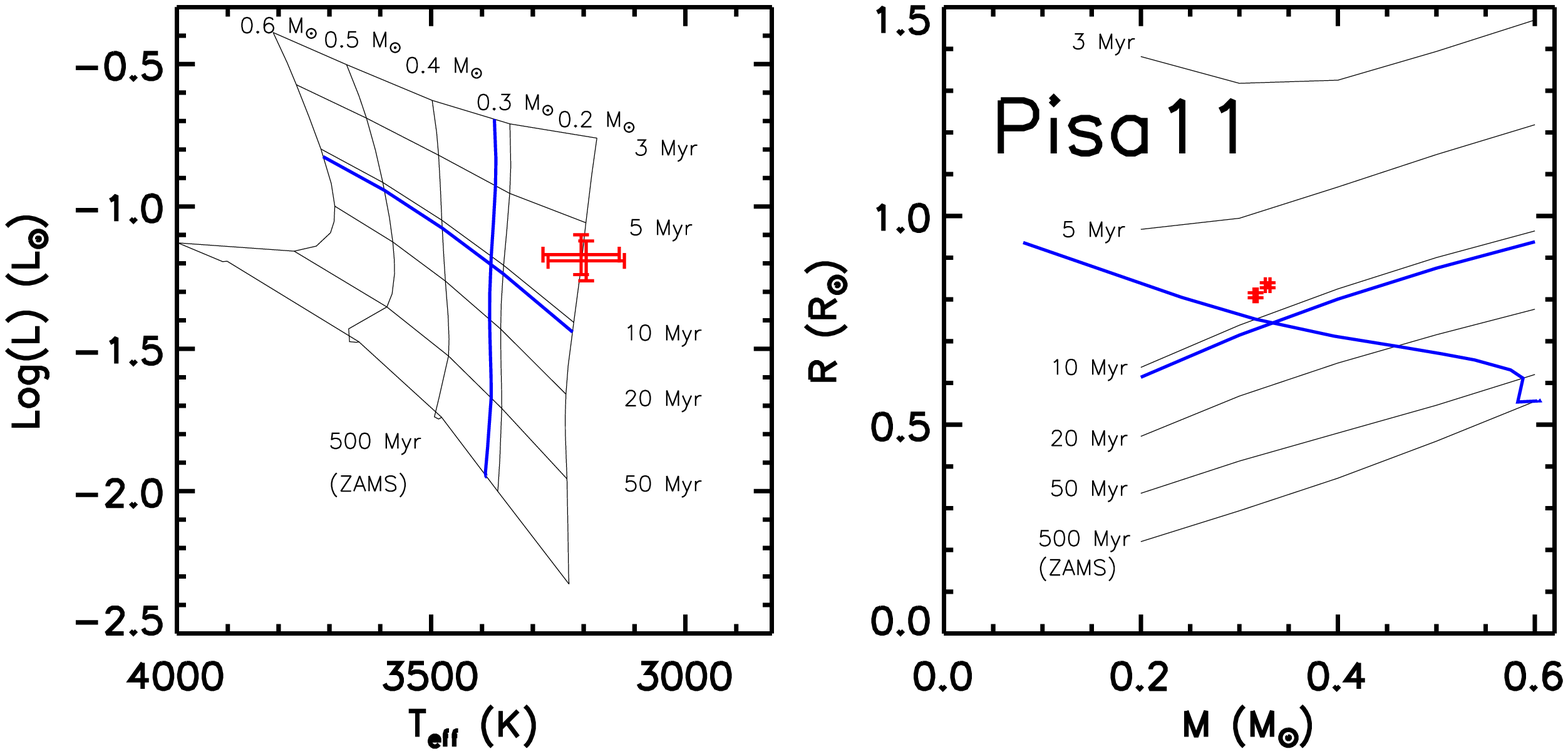}
 \caption{As in Figure~\ref{fig:bhac15}, but for the Pisa models \citep{Tognelli:2011fy}.} \label{fig:pisa}
  \end{figure*}
  
     \begin{figure*}
 \epsscale{1.12}
 \plotone{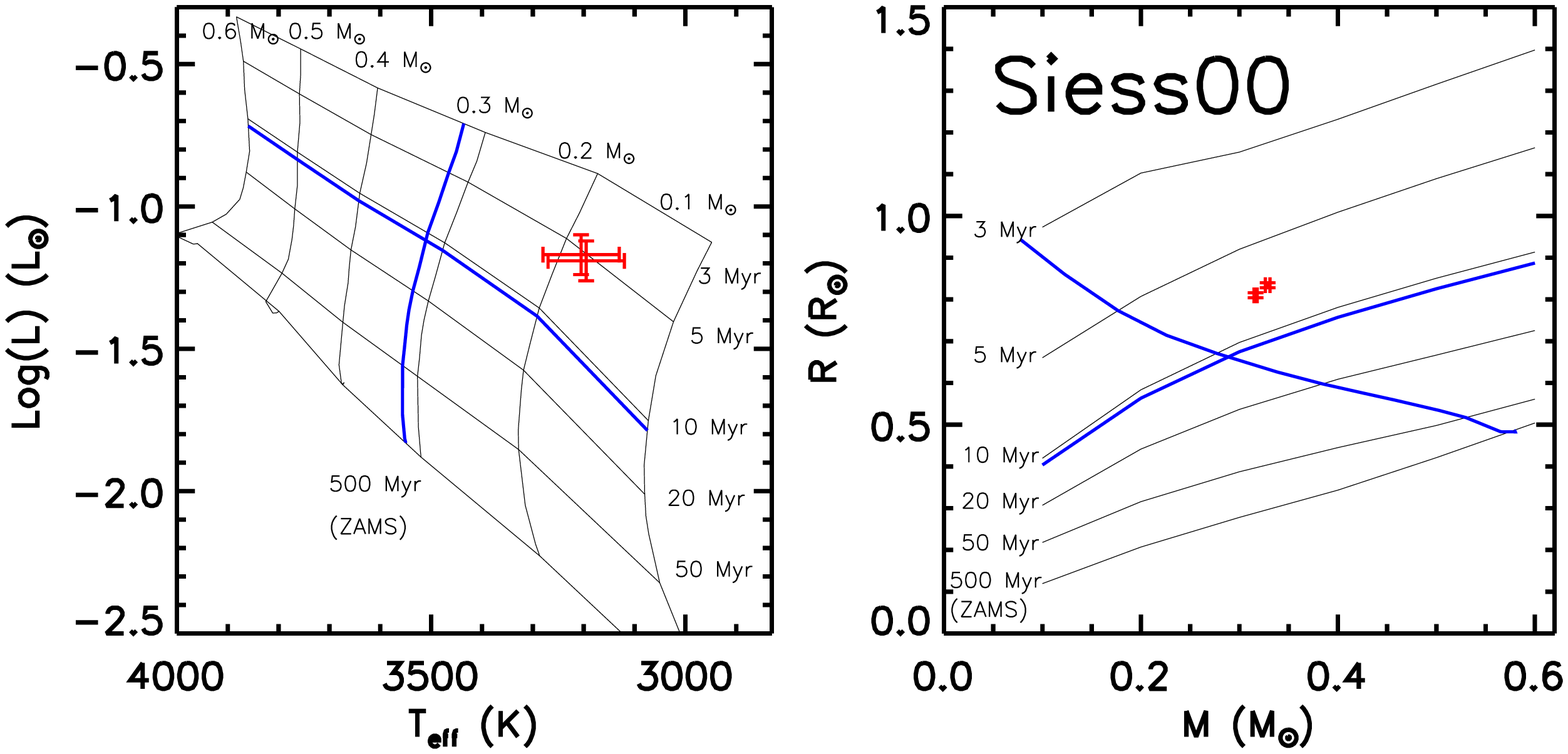}
 \caption{As in Figure~\ref{fig:bhac15}, but for the Siess models \citep{Siess:2000ce}.} \label{fig:siess}
  \end{figure*}

\begin{deluxetable}{lrrrr}
\tabletypesize{\footnotesize}
\tablewidth{0pt}
\tablecaption{Best-Fit Model Parameters}
\tablehead{
\colhead{Model Set} & \colhead{$\tau_{model}$} & \colhead{$M_{model}$} & \colhead{$\tau_{best}$} & \colhead{$\sigma$}
\\
\colhead{} & \colhead{(Myr)} & \colhead{($M_{\odot}$)} & \colhead{(Myr)}
}
\startdata
          BHAC15 & $ 4.8^{+2.8}_{-1.6}$ & $0.22^{+0.05}_{-0.03}$ &  9.7 & 2.4 \\
         Siess00 & $ 5.8^{+1.8}_{-1.2}$ & $0.20^{+0.03}_{-0.03}$ & 10.3 & 4.2 \\
            Pisa & $ 6.7^{+2.0}_{-1.0}$ & $0.23^{+0.04}_{-0.05}$ &  9.6 & 2.4 \\
          Padova & $13.1^{+4.1}_{-3.9}$ & $0.41^{+0.05}_{-0.05}$ &  6.6 & 1.7 \\
         DSEP    & $ 6.2^{+2.5}_{-1.5}$ & $0.23^{+0.05}_{-0.04}$ &  9.7 & 2.3 \\

\enddata
\tablecomments{The model masses are the values for each star, which are identical to well within the uncertainties in the HR diagram positions. The age $\tau_{best}$ is the age along each model set's isomass sequence that is closest to the system's true location in the HR diagram (Section 6.5); we also list the $\sigma$ level of the discrepancy. }
\end{deluxetable}

UScoCTIO 5 poses a strong challenge to the evolutionary models of young low-mass stars ($\tau \la $1--10 Myr; $M \sim $0.1--0.7 $M_{\odot}$), requiring predictive agreement with both the masses and the radii of its component stars, in addition to the luminosities and surface temperatures that are more commonly available for most young stars. This system also offers an intriguing older analog to the four previously-known eclipsing binary systems with young, low-mass components: 2M0535 (0.055 + 0.035 $M_{\odot}$; \citealt{Stassun:2006bf}), JW 380 (0.26 + 0.25 $M_{\odot}$; \citealt{Irwin:2007vg}), Par 1802 (0.39+0.39 $M_{\odot}$; \citealt{Gomez-Maqueo-Chew:2012dq}), and CoRoT 223992193 (0.67 + 0.50 $M_{\odot}$; \citealt{Gillen:2014nx}). Those systems represent even more extreme youth, with inferred ages of $\tau \sim$ 1--3 Myr for their host populations (the ONC and NGC 2264), but they also are complicated by the presence of circumstellar disks and approach the point where models are distinguished more by initial conditions than by stellar evolutionary processes, as well as having stronger tidal interactions due to their shorter orbital periods.

We specifically compare the properties of this touchstone system \citep{Mann:2015lr} to the predictions of five sets of evolutionary models: the BHAC15 models (Baraffe et al. 2015), the Dartmouth Stellar Evolution Program or DSEP models \citep{Dotter:2008qq,Feiden:2015jk}, the Padova/PARSEC models \citep{Bressan:2012hc,Chen:2014fk}, the Pisa models \citep{Tognelli:2011fy}, and the models of \citet{Siess:2000ce} that while older, remain in common usage. In the left panels of Figures 9--13, we show the (nearly identical) positions of the two components of UScoCTIO 5 in the $L$-$T_{\rm eff}$ HR diagram, as well as the isochronal and isomass sequences predicted by each of the four sets of models. In each figure we use the luminosity for UScoCTIO 5 calculated in Section 4.3 and the $T_{\rm eff}$ predicted by the L03 temperature scale. In Table 5, we list the corresponding model predictions for the mass and age of the system given its HR diagram position. To compute these predictions we adopted linear-uniform priors on both the mass (since the system falls near the peak of the IMF) and the age (implying a roughly constant star formation history in Upper Sco and the solar neighborhood). We also show the isomass sequence that the models predict for the observed dynamical masses in the system ($M \sim 0.32 M_{\odot}$) as well as the isochronal sequence predicted by observations of more massive Sco-Cen members ($\tau \sim 11$ Myr; \citealt{Pecaut:2012dp}) that continues to gain currency over the longstanding predicted age determined for low mass-stars ($\tau \sim 5$ Myr; \citealt{Preibisch:2002qt}).

We find that none of the model isomass sequences successfully predict the HR diagram position of UScoCTIO 5 to within the observational uncertainties, though the Padova models do predict an isomass sequence that agrees to within 2$\sigma$. This comparison is effectively univariate since the isomass sequences of fully convective stars (which evolve along the Hayashi track) are nearly vertical, and hence even the Padova models still disagree with the observed mass at $>$90\% in a one-sided test. Intriguingly, the sign and magnitude of the discrepancy is nearly identical between the BCAH15, DSEP, Pisa, and Siess models (where they under-predict the mass, as originally shown by \citealt{Reiners:2005wq}), whereas the Padova models overpredict the mass. The new Padova models \citep{Chen:2014fk} differ from the other model isochrones because they apply an empirical correction to the outer boundary correction (expressed as a $T_{eff}$ - $\tau$ relationship modification from the BT-Settl photospheres) in order to match observed mass-radius relationships for low-mass dwarfs. This appears to be an over-correction in the case of UScoCTIO 5, as would be expected if, for example, the reason for the correction is missing atmospheric opacities that are most important at high gravity. However, the fundamental limitations in measurement of $T_{\rm eff}$ (which depends on model atmosphere physics) and $L_{bol}$ (which depends on precise distances) will remain a limit on the utility of the HR diagram in testing evolutionary models.

In the right panels of Figures 9-13, we show the (also nearly identical) positions of the two components of UScoCTIO 5 in the mass-radius diagram, with the isochronal and iso-luminosity sequences of each model set for comparison. These observations pose a much more stringent test of the evolutionary models, since both the mass and the radius can be determined much more precisely in comparison to the dynamic range of the model predictions. As with the HR diagram, the two components appear very nearly coeval. The BCAH15, DSEP, Pisa, and Siess tracks all predict ages that are significantly younger than the newly-canonical age of $\tau \sim 11$ Myr inferred from the upper main sequence \citep{Pecaut:2012dp}. As with the HR diagram, the closest agreement is achieved by the Padova models, which almost exactly reproduce the expected age. 

However, we find that none of the model sets reproduce the luminosity at the give mass, with discrepancies that follow those of the HR diagrams. For BHAC15, DSEP, Pisa, and Siess, the models predict radii that are too small for the known luminosity and mass of the stars, indicating (from the Stefan-Boltzmann law) that the model temperatures are too high. Conversely, even though the Padova models predict the canonical age of $\tau \sim 11$ Myr for Upper Sco, the model radius for the known luminosity and mass is too high, indicating an underpredicted temperature. We therefore can demonstrate the same conclusions as for the HR diagrams (that there are discrepancies in $T_{\rm eff}$) without relying on the systematic uncertainties of directly measuring stellar $T_{\rm eff}$ from the observed spectral types.

Stars like UScoCTIO 5 remain fully convective throughout their evolution toward the ZAMS, and hence evolution on the pre-main sequence largely consists of dimming at constant temperature as the star contracts. The functional result is that luminosity is a first-order predictor of age, while temperature is a first-order predictor of mass. If a star does not fall on the appropriate isomass sequence for a model set, it therefore indicates either that the models are not predicting the correct value of $T_{\rm eff}$, or that the observational spectral types are not being mapped correctly to to the underlying $T_{\rm eff}$ of the stars' true atmospheres. If the former case is true, then it indicates a discrepancy in how the models handle energy transport. Either convection is less efficient in the interior \citep{Mullan:2001pm,Chabrier:2007cs,Gennaro:2012oq} or opacities are missing in the atmosphere (e.g., \citealt{Rajpurohit:2013qy}); either would result in a smaller radius and a hotter temperature in the models, as we see for all but the Padova set. If the temperature scale is not correct (for example, due to gravity-dependent changes in the appearance of major molecular bands), then the true stellar temperature could be hotter or colder than predicted.

\subsection{An Empirical Constraint on the Temperature Scale of Young Stars}

The model comparisons in Section 6.1 are predicated on a stellar effective temperature derived from the ad hoc young-star temperature scale of L03. This temperature scale was designed to bridge the large difference between dwarf and giant temperatures at a given spectral type, and specifically to make the GG Tau quadruple and the low-mass stellar sequence of IC 348 appear coeval when compared to the models of \citet{Baraffe:1998yo}. However, this temperature scale is still almost totally uncalibrated with observations, and there are now indications that the GG Tau multiple system could host additional components \citep{Di-Folco:2014uq}. Almost all young stars are too distant to measure interferometric sizes, preventing a direct measurement of stellar radii. The UScoCTIO 5 system (and other future discoveries) offer the intriguing alternative of measuring geometric radii from eclipsing binaries that are nominally model-independent.

As we describe in Section 4.3, we estimate that the two components of UScoCTIO 5 have a combined luminosity of $L_{bol} \sim 0.132 \pm 0.03 L_{\odot}$. Given effective temperatures for each component that are virtually identical ($\Delta T_{\rm eff} \la 10$ K), then the corresponding absolute temperatures can be derived purely from the sum of the radii and the Stefan-Boltzmann Law. We find that $T_{\rm eff} = 0.558 T_{\odot} = 3260^{+160}_{-200}$ K. This measurement is only discrepant from the values predicted for dwarfs or giants (3100 K or 3435 K) by $\pm$1 sigma, and hence does not yet provide a useful constraint on the system. However, the measurement is limited almost entirely by the uncertainty in the distance ($d = 145 \pm 15$ pc). This system falls well within the brightness range where \textit{Gaia} should deliver extremely precise parallaxes within 2 years; once those results are released (or if the distance can be refined in some other way), the system temperature can be described by:

\begin{eqnarray}
T_{\rm eff} = (3260^{+50}_{-30} K)(\frac{D}{145 pc})^2
\end{eqnarray}

\subsection{An Infrared Flux Method Distance}

As we have directly measured the radii of the two stars in linear units, and have bolometric luminosities and spectroscopically estimated $T_{\rm eff}$ values, we can estimate the angular diameters and therefore the distance to the system by simple trigonometry. As this is somewhat dependent on reddening, we can improve this distance measurement using the infrared flux method (e.g. \citealt{Casagrande:2010kq}). We computed a model K-band flux for the two stars interpolating the BT-SETTL models \citep{Allard:2011qf}, using the spectroscopically derived effective temperatures and extinction, the K filter profile from \citet{Cohen:2003yu}, converting $A_V$ to $A_K$ using \citet{Savage:1979jk}. The final distance computed is 144.4$\pm$6.6\,pc for the L03 temperature scale or 135.1$\pm$8.8\,pc for the \citet{Herczeg:2014oq} temperature scale, in good agreement with the assumed distance of 145\,pc to Upper Scorpius. The distance uncertainty is dominated by the uncertainty in the spectroscopic effective temperature.

\subsection{The Coevality of Young Binary Systems}

The apparent coevality of UScoCTIO 5 A+B poses a counterpoint to recent suggestions that (apparent) stellar ages could be essentially ``randomized'', which threatens the assumption underlying all studies of stellar evolution that treat star clusters as simple stellar populations. Apparent non-coevality has been seen at ages of $\tau \sim 1$ Myr for the eclipsing binary systems 2M0535 and Par 1802 \citep{Stassun:2007bb,Gomez-Maqueo-Chew:2012dq}, where precise characterization of the stellar parameters shows that the binary components do not lie on the same isochrones. Non-coevality also has been suggested more generally as a source of intrinsic luminosity spreads within binary systems and stellar populations, though tests of non-coevality require careful consideration of the non-linearity of isochrones and mass tracks, as noted by \citet{Gennaro:2012oq}. Two modes have been suggested as possible sources of this randomization, one tied to the assembly and one tied to the subsequent evolution. 

One phenomenon that could alter apparent stellar ages is through stochastic variations in the episodic accretion history of individual stars \citep{Baraffe:2010lr}, whereby a variable fraction of the accretion energy is radiated away during the accretion process and not deposited into the star. Stars that radiate away more accretion energy are left with smaller radii at a given mass and age. Close binaries are an imperfect test of the accretion hypothesis since the stars likely accrete from a circumbinary disk and hence should have correlated accretion histories. However, recent observations suggest that accretion could preferentially occur onto primary or secondary stars (e.g., \citealt{Jensen:2007zo}) depending on the specific angular momentum of the accreted material, and hence forced-coevality might not be assured.

The other phenomenon that could alter apparent ages is through variable magnetic field strengths \citep{Mullan:2001pm,Chabrier:2007cs}. Strong magnetic fields within the stellar interior should inhibit convection (preventing radial movement of charged particles), resulting in less efficient energy transport and a correspondingly larger radius and lower temperature for a given luminosity. Strong magnetic fields near the surface also should increase the starspot fraction, reducing the average surface temperature and hence again requiring a larger radius for a given luminosity. These effects could manifest as a correlation between fundamental properties (mass, radius, and temperature) and activity signatures such as UV or H$\alpha$ emission \citep{Stassun:2012lq,Stassun:2014fj}.

Given the very precise agreement in the apparent ages of UScoCTIO 5 A+B, combined with other internally coeval systems at younger ages (JW 380 in the ONC and CoRoT 223992193 in NGC 2264), it appears that ages are not significantly randomized for all young stars. We therefore suggest that the process forcing apparent non-coevality likely only occurs for a fraction of all stars. It has been demonstrated for visual binaries in Taurus ($\tau \sim 2$ Myr) that $\sim$2/3 of all pairs appear highly coeval ($\Delta \log(\tau) < 0.16$ dex or $<$40\%; \citealt{Kraus:2009fk}), indicating that $\le$20\% of all stars have apparent ages that differ substantially from the actual age.

\subsection{The Ages of UScoCTIO 5 and Upper Scorpius}

The age of the Upper Scorpius OB association, and by extension all young populations age-dated in a similar manner, is a topic of contention in the current literature. The ages of many populations have been set by the positions of low-mass stars (such as UScoCTIO 5) in the HR diagram, as compared to isochrones predicted by stellar evolutionary models. However, the ages of upper main sequence stars (such as the F stars in Upper Sco) and the location of the age-dependent lithium depletion boundary seem to predict ages that are older by a factor of two. The most visible debate has occurred for Upper Sco itself, with predictions of $\tau \sim 5$ Myr from low-mass stars \citep{Preibisch:2002qt,Slesnick:2006pi} and $\tau \sim 11$ Myr from the other methods \citep{Pecaut:2012dp}.

UScoCTIO 5 represents a fundamentally new datapoint for this debate. If the evolutionary models of low-mass stars are indeed predicting the correct age, then they should also predict the correct mass and radius for a given luminosity and spectral type. Conversely, if they are predicting incorrect ages because either the luminosity or temperature predictions are incorrect, then the (time-independent) masses or the (time-dependent) radii might not match. This comparison is especially useful for low-mass stars because they evolve along the fully-convective Hayashi track and hence fall nearly vertically in the HR diagram; $T_{\rm eff}$ corresponds mostly to mass, while $L_{bol}$ corresponds mostly to age.

As can be seen in Figures 9, 10, 12 and 13, and as we summarize in Table 5, the BHAC15, DSEP, Pisa, and Siess tracks do indeed predict a younger age for UScoCTIO 5 than the canonical value estimated for intermediate-mass stars. However, the model isomass sequences all predict that UScoCTIO 5 should have higher $T_{\rm eff}$ than is observed, generally approaching or exceeding the value that would be expected for an M4.5 giant (indicating that temperature scale changes alone might not solve this problem). The spectral line lists for low-mass stars are known to be incomplete, and hence the models must be missing opacities that would drive their predicted temperatures lower. If the $T_{\rm eff}$ predictions were shifted to bring the isomass sequences into agreement with the observations, given the same luminosity, then all three sets of models would indeed predict an age of $\tau \sim 11$ Myr; we list that best-fitting age along each model's isomass sequence in Table 5, along with the $\sigma$ level of the discrepancy. This is unlikely to completely solve the problem, though, since the models can't be shifted purely horizontally. Lower temperatures would result in less energy being radiated away, slowing the contraction and hence also modifying the radius and luminosity at a given age.

In contrast, Figure 11 and Table 5 show that the Padova models predict an older age from the HR diagram of UScoCTIO 5, while under-predicting the temperature for stars that fall along its isomass sequence. If the model isomass sequence were shifted in $T_{\rm eff}$/$L_{bol}$ space to match the observed $T_{\rm eff}$, then the inferred age would be younger, but still consistent with $\tau \sim 11$ Myr to within $\sim 1 \sigma$. Some authors do indeed predict that the young-star temperature scale should fall at even cooler temperatures (e.g., \citealt{Herczeg:2014oq}), which would bring UScoCTIO 5's SpT-derived $T_{\rm eff}$ into excellent agreement with the Padova grid's existing predictions for the $T_{\rm eff}$ and $L_{bol}$ of a $\sim$10 Myr, $\sim0.3 M_{\odot}$ star.

However, as the discussion above demonstrates, the uncertainties in UScoCTIO 5's HR diagram position are dominated by the $\pm$75 K observational uncertainty and the $\pm$100 K systematic uncertainty due to the unknown gravity dependence of the temperature scale for young M-dwarf stars. A more robust comparison can be made in the mass-radius plane. Both quantities for the system can be determined with greater precision compared to the dynamic range of the model predictions, as well as without systematic uncertainty. As can be seen in the right-hand panels of Figures 9-13, the BHAC15, DSEP, Pisa, and Siess models do indeed predict ages that are significantly younger than 10 Myr, though none fall as young as 5 Myr, while the Padova models predict the older age seen for intermediate-mass stars. 

The robustness of these predictions, and the direction of any discrepancy, can be tested by comparison to the iso-luminosity line for UScoCTIO 5. The intersection of the iso-luminosity line and the 11 Myr isochrone falls very nearly at the mass of UScoCTIO 5 for the BHAC15, DSEP, and Pisa models, at a moderately lower mass for the Siess tracks, and at a much higher mass for the Padova tracks. These results support the trend seen for the HR diagram - if the $T_{\rm eff}$ and radius predicted by the models are modified to match the isomass lines, then most of the tracks would indeed predict an age of $\tau \sim 11$ Myr. We therefore conclude that while none of the model sets predict all system parameters perfectly, the likely form of the discrepancy supports the ongoing rescaling of pre-main sequence stellar ages in favor of older values.

\section{Summary}

We have presented the discovery that UScoCTIO 5, a known low-mass spectroscopic binary ($P = 34$ days, $M \sim 0.64 M_{\odot}$) in the Upper Scorpius star-forming region ($\tau \sim 10$ Myr), is an eclipsing system suitable for determination of precise stellar masses and radii. Based on the stellar properties ($M_A = 0.329 \pm 0.002 M_{\odot}$, $R_A = 0.834 \pm 0.006 R_{\odot}$, $M_B = 0.317 \pm 0.002 M_{\odot}$, $R_B = 0.810 \pm 0.006 R_{\odot}$), we conclude that:

\begin{enumerate}

\item There are systematic errors in the calibration of pre-main sequence evolutionary models. The BHAC15, DSEP, Pisa, and Siess models overpredict the $T_{\rm eff}$ of young stars for a given mass by $\sim$200 K, or equivalently underpredict the masses of young stars for the given $T_{\rm eff}$ by $\sim$50\%. The Padova models are a slightly better match, but are discrepant in the opposite direction, underpredicting $T_{\rm eff}$ by $\sim$100 K or overpredicting mass by $\sim$25\%. The discrepancies remain in the mass-radius-luminosity space, suggesting that the discrepancies likely represent intrinsic calibration issues rather than an uncertain temperature scale.

\item Our geometric measurement of $T_{\rm eff}$ (derived from radius and luminosity) broadly agrees with the temperature scale for young stars, but the uncertainty will remain too large to refine the temperature scale until the luminosity can be measured more precisely with a \textit{Gaia} distance.

\item UScoCTIO 5 appears highly coeval, bringing the count among low-mass EBs to three apparently coeval pairs and two apparently non-coeval pairs. The inferred fraction of stars with spurious ages ($\sim$20\%) is consistent with the number seen for wider visual binaries, suggesting that processes which randomize apparent stellar ages do occur, but in a minority of cases.

\item Taking into account the dimensions within which the models appear to be miscalibrated, we find that the age of UScoCTIO 5 appears more consistent with the older age of Upper Scorpius that has recently gained canonical status: $\tau \sim$ 11 Myr.

\end{enumerate}

\acknowledgements

We thank T. Dupuy, K. Larson, G. Herczeg, A. Dotter, G. Feiden, and M. Bessel for helpful discussions, and J. Bento and R. Kuruwita for assisting with the WiFeS observations. We also thank the referee, Eric Mamajek, for providing an insightful and prompt critique that improved the quality of this work.

This research was partially supported by an appointment to the NASA Postdoctoral Program at the Ames Research Center, administered by Oak Ridge Associated Universities through a contract with NASA.

This paper includes data collected by the K2 mission. Funding for the K2 mission is provided by the NASA Science Mission directorate. The K2 data presented in this paper were obtained from the Mikulski Archive for Space Telescopes (MAST). This research also has made use of the Keck Observatory Archive (KOA), which is operated by the W. M. Keck Observatory and the NASA Exoplanet Science Institute (NExScI), under contract with the National Aeronautics and Space Administration. The archival Keck/HIRES observations herein were obtained at the W.M. Keck Observatory by PIs G. Basri, W. Sargent, and J. Kuhn. Keck is operated as a scientific partnership among the California Institute of Technology, the University of California and the National Aeronautics and Space Administration. The Observatory was made possible by the generous financial support of the W.M. Keck Foundation. 

The authors wish to recognize and acknowledge the very significant cultural role and reverence that the summit of Mauna Kea has always had within the indigenous Hawaiian community.  The archival PIs were most fortunate to have the opportunity to conduct observations from this mountain.

\appendix

The light curves of eclipsing binaries are traditionally predicted using dedicated software packages (such as the Wilson-Devinney code; \citealt{Wilson:1971ig}). These packages include a large number of physical effects that can be relevant for some systems, such as tidal distortion, reflected light, and star spots. However, the architecture needed to model these effects results in a code that requires seconds to produce a single light curve, even when the features are turned off and even though it runs in a compiled language (Fortran). This runtime makes MCMC implementations onerous to execute, since long chains and multiple walkers are needed to converge and become well-mixed in the high-dimensional parameter space of eclipsing binary parameters. 

We therefore present here a modification of the formalism of \citet{Mandel:2002ai} that is commonly used to model transiting extrasolar planets. We specifically use their formalism (and their IDL implementations)\footnote{\url{http://www.astro.washington.edu/users/agol/transit.html}} to predict the flux decrement from whichever star is being occulted, and then include an additional parameter (the surface brightness ratio) that, in combination with the radius ratio, can be used to add the light of the occulting star. 

Consider two stars in an eclipsing binary, hereafter P and S, that have radii $R_p$ and $R_s$ (with ratio $r = R_s/R_p$) and surface brightnesses $S_p$ and $S_s$ at some arbitrary wavelength (with ratio $s = S_s/S_p$). The surface brightnesses will depend on $T_{\rm eff,p}$ and $T_{\rm eff,s}$, but for cool stars with complicated spectra, this dependence might not be straightforward. For simplicity, this algorithm fits for the ratio of surface brightnesses $s$, but does not attempt to calculate the corresponding temperatures.

In the case where neither star is eclipsed, denoted time $t=0$, the observed flux is:

\[F_{tot}(0) = F_p(0) + F_s(0) = \pi \frac{R_p^2}{D^2} S_p + \pi \frac{R_s^2}{D^2} S_s \]

Potentially also including a limb darkening term; in our case this term is identical for both stellar components, and hence it factors out. If the primary star is eclipsed at a time $t_1$ with an impact parameter $z_1(t_1)$, then the Mandel-Agol algorithm can be called to calculate the fractional flux deficit with respect to the primary star's total flux, $\mu_p(z_1(t_1)) = \frac{F_p(t_1)}{F_p(0)}$. The other star is still contributing $F_s(t_1)=F_s(0)$, and hence the total flux observed is:

\[F_{tot}(t_1) = F_p(t_1) + F_s(t_1) = \mu_p(z_1(t_1))F_p(0) + F_s(0) = \mu_p(z_1(t_1)) \pi \frac{R_p^2}{D^2} S_p + \pi \frac{R_s^2}{D^2} S_s\]

If we normalize this measurement by the out-of-eclipse total flux, then we can recast the observation as:

\[\frac{F_{tot}(t_1)}{F_{tot}(0)} = \frac{\mu_p(z_1(t_1)) R_p^2 S_p + R_s^2 S_s}{R_p^2 S_p + R_s^2 S_s} = \frac{\mu_p(z_1(t_1))  + r^2 s}{1 + r^2 s} \]

Similarly at a time $t_2$ during the secondary eclipse, then:

\[F_{tot}(t_2) = F_p(t_2) + F_s(t_2) = F_p(0) + \mu_2(z_2(t_2))F_s(0) = \pi \frac{R_p^2}{D^2} S_p + \mu_s(z_2(t_2)) \pi \frac{R_s^2}{D^2} S_s\]

and

\[\frac{F_{tot}(t_2)}{F_{tot}(0)} = \frac{R_p^2 S_p + \mu_s(t_2) R_s^2 S_s}{R_p^2 S_p + R_s^2 S_s} = \frac{1 + \mu_s(z_2(t_2)) r^2 s}{1 + r^2 s} \]

The only input required to calculate $\mu(z(t))$ is the impact parameter between the two stars as a function of the occulted star's radius, which can be calculated analytically from the geometry of the orbit for any given orbital phase, and hence this formalism can be used to calculate the time-dependent fractional flux (with respect to non-eclipse epochs) observed for an eclipsing binary system over the course of its eclipses. Even implemented in an interpreted language (IDL) that is inherently much slower, this routine produces light curves 10--100 times faster than when using the latest Wilson-Devinney code\footnote{\url{ftp://ftp.astro.ufl.edu/pub/wilson/}} and therefore provides a suitable fast approximation when the more powerful features of Wilson-Devinney are not required.

\bibliographystyle{apj.bst}
\bibliography{ms.bib}

\end{document}